\documentclass[fleqn,usenatbib]{mnras}

\usepackage{newtxtext,newtxmath}

\usepackage[T1]{fontenc}

\DeclareRobustCommand{\VAN}[3]{#2}
\let\VANthebibliography\thebibliography
\def\thebibliography{\DeclareRobustCommand{\VAN}[3]{##3}\VANthebibliography}

\usepackage{graphicx}	
\usepackage{amsmath}	

\usepackage{multirow}
\usepackage[dvipsnames]{xcolor}
\usepackage{float}

\newcommand{\bham}{School of Physics and Astronomy, University of Birmingham, Birmingham B15 2TT, UK}
\newcommand{\igw}{Institute for Gravitational Wave Astronomy, University of Birmingham, Birmingham B15 2TT, UK}

\newcommand{\qub}{Astrophysics Research Centre, School of Mathematics and Physics, Queens University Belfast, Belfast BT7 1NN, UK}

\title[Bulge masses of TDE hosts]{Evidence for a steeper SMBH--Bulge mass relationship extended to low masses using TDE host galaxies}

\author[P.~Ramsden et al]{{Paige Ramsden$^{1,2}$}\thanks{Contact e-mail: \href{mailto:pxr754@alumni.bham.ac.uk}{pramsden@star.sr.bham.ac.uk}}, Matt Nicholl$^{2}$, Sean L.~McGee$^{1, 3}$, and Andrew Mummery$^{4}$
\\
$^{1}$\bham\\
$^{2}$\qub\\
$^{3}$\igw\\
$^{4}${Oxford Theoretical Physics, Beecroft Building, Clarendon Laboratory, Parks Road, Oxford, OX1 3PU, United Kingdom}
}

\usepackage{lscape}

\date{Accepted XXX. Received YYY; in original form ZZZ}

\pubyear{2021}

\begin{document}
\label{firstpage}
\pagerange{\pageref{firstpage}--\pageref{lastpage}}
\maketitle

\begin{abstract}

Tidal disruption events (TDEs) are excellent tools for probing low mass supermassive black holes (SMBHs) that may otherwise remain undetected. Here, we present an extended SMBH--Bulge mass scaling relationship using these lower mass TDE black holes and their host galaxies. Bulge masses are derived using \textsc{Prospector} fits to UV-MIR spectral energy distributions for the hosts of 40 TDEs that have a detected late-time UV/optical plateau emission, from which a SMBH mass is derived. Overall, we find that TDE plateaus are a successful method for probing BH scaling relations. We combine the observed TDE sample with a higher mass SMBH sample and extend the known relationship, recovering a steeper slope ($m = 1.34 \pm 0.03$) than current literature estimates, which focus on the high mass regime. For the TDE only sample, we measure an equally significant but shallower relationship with a power-law slope of $m = 1.17 \pm 0.10$ and significance $<0.001$. Forward modelling is used to determine whether known selection effects can explain both the comparatively flatter TDE only relation and the overall steepening across the full SMBH mass range. We find that the flattening at TDE masses can be accounted for, however the steepening can not. It appears that if a single slope extends for the whole BH mass range, it must be steeper to include the TDE population. 

\end{abstract}




\section{Introduction}
\label{s:intro}

With supermassive black holes (SMBHs) residing in the nuclei of most galaxies, tidal disruption events (TDEs) occur when an unfortunate star is perturbed into the galactic centre. The star is torn apart when the tidal forces it experiences overcome its own self-gravity \citep{hills, rees}. If the orbital pericentre is close to the critical `tidal radius', and that tidal radius lies outside of the black hole's event horizon, it is theorised that roughly half of the stellar debris becomes bound to the SMBH and the resulting accretion of this bound matter produces an observable flare \citep{lacey}. If instead the star intersects the horizon before disruption (i.e. the tidal radius is inside the gravitational radius), it will be swallowed whole and no flare will be observed. This criterion that the tidal radius must lie outside the gravitational radius, introduces a critical SMBH mass, the Hills mass, below which TDEs can be observed \citep{hills}.

For a main-sequence star of mass $M_*$$\sim$1${\rm M}_\odot$, and radius $R_*$$\sim$1${\rm R}_\odot$, the maximum mass for a SMBH that can produce an observable TDE is $\lesssim 10^8 {\rm M}_{\odot}$. Note that as well as depending strongly on SMBH mass, this critical Hills mass is also impacted by spin. For a maximally rotating SMBH the Hills limit softens, enabling observations of TDEs around somewhat more massive SMBHs \citep{mummery2023b}. By nature, TDEs are excellent probes for the lower mass-regime of black holes, successfully identifying systems that would otherwise lie undetected. These events also provide the opportunity to study accretion physics, the evolution of SMBHs over cosmic time, and the link between SMBHs and their host environments.    

In comparison to other transient events, TDEs remain intrinsically rare. However, they are now routinely found by wide-field, time-domain surveys, with an estimated occurrence rate of one every 10$^{4-5}$ years per galaxy \citep{yao23,sazonov}. While similar in brightness to some supernovae (SNe), with a characteristic luminosity of $\sim 10^{44}$ erg s$^{-1}$, they have been detected only in galactic nuclei and have a distinct lack of colour evolution; a longer rise and fade time than most SNe; and a slow decay often approximated as a power-law or exponential over time \citep{phinney,van_class5}. Certain spectroscopic features also distinguish TDEs from other transients, including a strong blue continuum with colour temperature of 10,000 - 50,000 K and broad emission lines \citep{hung}.

Studies of BH demographics \citep{kormendy} have shown that SMBH properties correlate strongly with those of the galaxies that host them, and relationships linking SMBH mass with galaxy bulge mass and velocity dispersion were derived \citep{haring,Gebhardt2000, ferrarese,magorrian}. Constraining the slopes of these relationships enables the estimation of SMBH masses for large samples, furthering our understanding of early structure formation and the evolution of SMBHs \citep{Gultekin2009,mcconnell}. Particularly at low masses, the slope of BH scaling relations may reflect the physical mechanisms by which SMBHs grow \citep{greene20}. However, as most direct estimates of SMBH masses (outside of a handful of very nearby galaxies) are made by reverberation mapping in active galactic nuclei, relationships are often derived using samples dominated by SMBHs of mass $\gtrsim10^7$\,M$_\odot$ \citep[e.g.][]{kormendy,Bentz2013,mcconnell}. 

In addition to reverberation mapping, BH demographics can be explored using alternative techniques such as AGN broad line measurements \citep{reines15}, megamaser disks \citep{greene16}, and stellar kinematics \citep{krajnovi, thater,nguyen}. These methods can probe lower mass black hole samples and allow for a broader analysis of the correlation between SMBH and stellar mass \citep{reines15,greene20}. Often, when dealing with intermediate-mass black holes or low-mass host galaxies, studies consider the relation in terms of total stellar mass rather than bulge mass due to challenges in resolving bulges through ground-based imaging. However, the tightest correlations appear to be between the SMBH and bulge properties \citep{kormendy}.

The limiting Hills mass restricts TDEs to the lower end of the SMBH mass function. This has been confirmed by previous TDE host galaxy studies, which estimate SMBH masses using established galaxy scaling relations, finding them to be in the range $10^{4} \lesssim M_{\text{BH}} \lesssim 10^{8}$\,M$_\odot$ \citep{wevers19,yao23}. These estimates assume that the scaling relations derived at high galaxy mass also apply in the lower mass hosts of TDEs. However, if SMBH masses can be inferred directly using TDE emission, rather than properties of the host, TDEs could provide the perfect opportunity to test this assumption. With a large enough sample, TDEs could even be used to improve the calibration of scaling relations at the low mass end, and to test whether this differs from the high-mass regime. If we find that properties of TDE-selected BHs do couple with those of their hosts, the TDE BH demographic can be estimated for the predicted influx of detections using host galaxy photometry alone. 

Previous efforts to constrain scaling relations using TDEs have shown mixed results. The powering mechanism behind the early time optical/UV emission from TDEs is still debated. Current models assume that the observed flare is produced either by the reprocessing of X-ray emission from an accretion disk \citep{roth16} or shocks from debris stream collisions or during disk formation \citep{piran}. Early studies based on a small number of sources and using different codes such as \textsc{mosfit} (where luminosity tracks fallback/accretion rate; \cite{mockler}) and \textsc{TDEmass} (luminosity from stream collisions; \cite{ryu}) found that BH masses derived from TDE light curve models were consistent with host scaling relations. However, \cite{hammerstein2022} recently applied the two different models to a population study of TDEs discovered by ZTF. No correlation was found between the SMBH masses derived from these models and the total stellar masses of their host galaxies. Additionally, using \textsc{mosfit},  \cite{ramsden} found only a weak positive correlation between SMBH and host galaxy bulge mass, with the gradient only 2-$\sigma$ away from being negative. While the early time emission mechanism from TDEs is contested, at late times it is well understood as emission from an optically thick accretion disk, which, due to a combination of cooling and expanding, undergoes a plateau in optical and UV luminosity \citep{mummery20}. Importantly, \cite{mummery2023} find that the plateau luminosity correlates positively with the SMBH mass, with $L_{\text{plat}} \propto M_{\text{BH}}^{2/3}$, and confirm that the simulations of disk emission reproduce the observed TDEs. 

Thousands of TDEs are predicted to be detected per year along with exceptionally deep images of their host galaxies, thanks to the upcoming Vera Rubin Observatory and its Legacy Survey of Space and Time (LSST; \citealt{lsst}) \citep{bricman, yao23}. Deriving the rate of TDEs as a function of BH mass for this upcoming sample would be invaluable in developing our understanding of both TDEs and low mass SMBHs. However, in order to obtain the most precise BH mass estimates, spectroscopic measurements of the orbital velocity distribution of stars and/or gas in the galactic nucleus are required \citep{kormendy}. For the large sample of future TDEs, it will not be possible to obtain spectroscopic BH mass estimates for all TDEs. Even direct modelling of TDE light-curves in order to derive dynamical quantities such as SMBH mass, as in previous work \citep{nicholl2022, mockler}, requires multi-wavelength (specifically UV) follow-up of each event which again will be impractical for such a large number of TDEs. However, accurate host galaxy scaling relations will enable the derivation of BH masses for all TDEs with deep LSST host imaging, even if no plateau is detected.

In this work, we model host galaxy photometry with \textsc{Prospector}, a stellar population synthesis code, and derive the bulge masses for 41 TDEs. SMBH masses were separately estimated in \cite{mummery2023}. Combining the BH and galaxy bulge masses for our TDE sample with a higher mass SMBH sample \citep{kormendy}, we attempt to extend the slope of the SMBH--Bulge to $10^5 M_{\odot} \lesssim M_{\text{BH}} \lesssim 10^{10} M_{\odot}$. A strong positive correlation is recovered. Comparatively, we find a still significant but shallower relationship within the TDE only sample. This may be the result of a number of selection effects, such as the bias TDEs have towards low-mass SMBHs due to the critical Hills mass and the Malmquist luminosity bias optical surveys have towards more luminous sources (and therefore against lower mass candidates). 

In section \ref{seds}, we present the sample of TDE candidates and their host galaxies and describe the \textsc{Prospector} model used for modelling host galaxy light. In section \ref{bh-bulge}, we analyse the BH--bulge mass relationship for the TDE sample and report the results and significance. The TDE sample is then combined with that of the high mass regime \citep{kormendy} and the SMBH--Bulge mass relationship is derived for the whole range of SMBH masses. The relation we find appears steeper than the literature estimates of scaling relations using mostly  higher mass BHs. In section \ref{simulation} we compare the observed scaling relation to a forward model, and show that known selection effects cannot explain the steeper relationship. Future outlook and conclusions are presented in sections \ref{future} and \ref{conclusion} respectively.

\hypersetup{citecolor=black}
\begin{table*}\label{obs_tdes}
\small
\caption{\small Sample of 40 plateau TDEs with location, redshift and the derived host galaxy bulge mass.}
        \begin{center}
\renewcommand{\arraystretch}{1.25}
\begin{tabular}{l l l l l l l} 
 \hline\hline
 IAU Name/Discovery Name & RA & Decl. & Redshift (z)  & $\log_{10}(M_{\rm bulge} / M_{\odot})$ & $\log_{10}(M_{\rm BH} / M_{\odot})$ $^a$ \\ 
 \hline
 ASASSN-14ae & 11:08:40.120 & +34:05:52.23 & 0.043 & $9.65^{+0.15}_{-0.10}$ & $6.13^{+0.55}_{-0.42}$ \\
 ASASSN-14li & 12:48:15.230 & +17:46:26.44 & 0.021 & $9.65^{+0.09}_{-0.12}$ & $6.14^{+0.55}_{-0.42}$ \\
 ASASSN-15oi & 20:39:09.180 & - 30:45:20.10 & 0.020 & $8.76^{+0.18}_{-0.19}$ & $5.86^{+0.55}_{-0.44}$ \\
 AT2018hco & 01:07:33.635 & +23:28:34.28 & 0.090 & $9.53^{+0.18}_{-0.21}$ & $7.08^{+0.44}_{-0.40}$ \\
 AT2018hyz & 10:06:50.871 & +01:41:34.08 & 0.046 & $9.40^{+0.25}_{-0.29}$ & $6.91^{+0.49}_{-0.41}$ \\
 AT2018jbv & 13:10:45.558 & +08:34:04.28 & 0.340 & $9.89^{+0.19}_{-0.21}$ & $8.08^{+0.34}_{-0.35}$ \\
 AT2018zr & 07:56:54.530 & +34:15:43.61 & 0.071 & $9.40^{+0.21}_{-0.23}$ & $7.01^{+0.47}_{-0.40}$ \\
 AT2019azh & 08:13:16.945 & +22:38:54.03 & 0.022 & $9.77^{+0.09}_{-0.05}$ & $6.43^{+0.55}_{-0.41}$ \\
 AT2019bhf & 15:09:15.975 & +16:14:22.52 & 0.121 & $10.08^{+0.12}_{-0.12}$ & $6.59^{+0.54}_{-0.41}$ \\
 AT2019cho & 12:55:09.210 & +49:31:09.93 & 0.193 & $9.59^{+0.22}_{-0.25}$ & $6.58^{+0.54}_{-0.41}$ \\
 AT2019cmw & 18:48:39.479 & +51:00:48.73 & 0.518 & $10.80^{+0.15}_{-0.18}$ & $7.80^{+0.35}_{-0.35}$ \\
 AT2019dsg & 20:57:02.974 & +14:12:15.86 & 0.051 & $9.81^{+0.22}_{-0.31}$ & $6.88^{+0.49}_{-0.41}$ \\
 AT2019ehz & 14:09:41.880 & +55:29:28.10 & 0.074 & $9.36^{+0.17}_{-0.20}$ & $5.94^{+0.56}_{-0.43}$ \\
 AT2019meg & 18:45:16.180 & +44:26:19.21 & 0.152 & $9.66^{+0.20}_{-0.24}$ & $6.02^{+0.56}_{-0.43}$ \\
 AT2019qiz & 04:46:37.880 & - 10:13:34.90 & 0.015 & $8.97^{+0.10}_{-0.11}$ & $5.42^{+0.58}_{-0.45}$ \\
 AT2020acka & 15:55:01.935 & +16:18:16.17 & 0.353 & $10.80^{+0.17}_{-0.19}$ & $8.13^{+0.32}_{-0.40}$ \\
 AT2021axu & 11:46:36.356 & +30:05:07.43 & 0.190 & $9.79^{+0.17}_{-0.22}$ & $7.54^{+0.32}_{-0.36}$ \\
 AT2020neh & 15:21:20.090 & +14:04:10.52 & 0.062 & $9.52^{+0.11}_{-0.11}$ & $5.47^{+0.57}_{-0.46}$ \\
 AT2020opy & 15:56:25.728 & +23:22:21.15 & 0.159 & $10.18^{+0.12}_{-0.10}$ & $7.24^{+0.39}_{-0.39}$ \\
 AT2020qhs & 02:17:53.970 & - 09:36:50.85 & 0.345 & $10.54^{+0.22}_{-0.27}$ & $8.08^{+0.34}_{-0.35}$ \\
 AT2020vwl & 15:30:37.800 & +26:58:56.89 & 0.035 & $9.31^{+0.17}_{-0.19}$ & $6.09^{+0.55}_{-0.43}$ \\
 AT2020wey & 09:05:25.880 & +61:48:09.18 & 0.027 & $9.27^{+0.15}_{-0.13}$ & $5.07^{+0.62}_{-0.46}$ \\
 AT2020ysg & 11:25:26.022 & +27:26:26.16 & 0.277 & $10.23^{+0.18}_{-0.20}$ & $7.94^{+0.35}_{-0.36}$ \\
 AT2020yue & 11:00:00.329 & +21:06:45.86 & 0.204 & $10.37^{+0.15}_{-0.16}$ & $7.61^{+0.32}_{-0.36}$ \\
 AT2020zso & 22:22:17.130 & - 07:15:59.08 & 0.057 & $9.64^{+0.11}_{-0.09}$ & $5.64^{+0.57}_{-0.45}$ \\
 AT2021crk & 11:45:06.907 & +18:32:25.32 & 0.155 & $9.64^{+0.20}_{-0.24}$ & $6.56^{+0.54}_{-0.41}$ \\
 AT2021ehb & 03:07:47.806 & +40:18:40.57 & 0.018 & $9.26^{+0.24}_{-0.33}$ & $6.76^{+0.52}_{-0.41}$ \\
 AT2021gje & 16:50:07.332 & +34:49:07.82 & 0.358 & $10.99^{+0.13}_{-0.12}$ & $7.83^{0.33}_{0.34}$ \\
 AT2021mhg & 00:19:42.900 & +29:19:00.70 & 0.073 & $9.47^{+0.17}_{-0.19}$ & $6.52^{+0.55}_{-0.41}$ \\
 AT2021nwa & 15:53:51.279 & +55:35:19.67 & 0.047 & $9.70^{+0.15}_{-0.16}$ & $6.83^{+0.51}_{-0.40}$ \\
 AT2021sdu & 01:11:23.924 & +50:34:29.67 & 0.059 & $9.52^{+0.28}_{-0.38}$ & $7.21^{+0.39}_{-0.39}$ \\
 AT2021uqv & 00:32:39.880 & +22:32:56.04 & 0.106 & $10.01^{+0.12}_{-0.10}$ & $7.59^{+0.32}_{-0.35}$ \\
 iPTF-15af & 08:48:28.130 & +22:03:33.40 & 0.079 & $9.93^{+0.14}_{-0.12}$ & $6.10^{+0.55}_{-0.43}$ \\
 iPTF-16axa & 17:03:34.340 & +30:35:36.60 & 0.108 & $9.78^{+0.17}_{-0.21}$ & $6.68^{+0.53}_{-0.41}$ \\
 iPTF-16fnl & 00:29:57.010 & +32:53:37.24 & 0.016 & $9.39^{+0.13}_{-0.10}$ & $5.91^{+0.56}_{-0.44}$ \\
 PS1-10jh & 16:09:28.280 & +53:40:23.99 & 0.170 & $9.42^{+0.19}_{-0.23}$ & $6.63^{+0.53}_{-0.40}$ \\
 PTF-09djl & 16:33:55.970 & +30:14:16.65 & 0.184 & $9.96^{+0.20}_{-0.23}$ & $6.98^{+0.47}_{-0.41}$ \\
 PTF-09ge & 14:57:03.180 & +49:36:40.97 & 0.064 & $9.91^{+0.10}_{-0.12}$ & $6.36^{+0.55}_{-0.42}$ \\
 SDSS-TDE1 & 23:42:01.410 & +01:06:29.30 & 0.136 & $9.87^{+0.19}_{-0.23}$ & $6.85^{+0.50}_{-0.40}$ \\
 SDSS-TDE2 & 23:23:48.620 & - 01:08:10.34 & 0.252 & $10.10^{+0.27}_{-0.29}$ & $7.48^{+0.33}_{-0.36}$ \\
 \hline\hline
\end{tabular}\\
\hypersetup{citecolor=blue}
$^a$ From \citet{mummery2023}
\label{t:sample}
\end{center}
\end{table*}

\hypersetup{citecolor=blue}
\section{Spectral Energy Distribution Fitting}
\label{seds}

\subsection{Data Selection}

Our sample of TDE hosts is based on the work of \cite{mummery2023}, such that we can pair our \textsc{Prospector} TDE host galaxy masses to SMBH mass measurements derived using late-time optical/UV emission. This spans approximately 10 years and includes all optical TDEs detected up to mid-2019 \citep{vanvelzen2020}, as well as TDEs detected in the first half of the ZTF survey \citep{hammerstein2022} and second half up to late-2021 \citep{yao23}. Sources with detections at least 1 year post peak were selected. Our sample is then further restricted to TDEs with plateau detections (49 out of a total of 63 sources). Considering the predicted decline rate $t^{-5/3}$, a confirmed plateau detection requires a luminosity measurement in excess of the power-law decay model at the last observation of the event. Final cuts are made based on the available broadband photometry for each host galaxy. At minimum, the derivation of total stellar mass requires multi-colour optical imaging of the galaxy from wide-field surveys such as PanSTARRS \citep{panstarrs} or SDSS \citep{sdss}. Nine candidates did not pass this threshold, so the final sample includes 40 plateau TDEs and their host galaxies, presented in Table \ref{obs_tdes}.

We construct spectral energy distributions (SEDs) for each of the TDE host galaxies using broadband photometry. An automatic pipeline was developed for efficient collection of this data, \textsc{Galfetch}. The code queries: SDSS, PanSTARRS and DECAM \citep{decam} for optical data in $u$, $g$, $r$, $i$, $z$, $y$\footnote{$u$ bands from SDSS (and DECAM where possible), $y$ bands from PanSTARRS (and DECAM where possible).}; 2MASS, for near infrared (NIR; \cite{2mass}); unWISE, for mid-infrared (MIR; \cite{unwise}); GALEX, for UV \citep{galex}. Data collected from 2MASS was done via the NASA/IPAC Infrared Science Archive, whilst unWISE data was pulled from DESI Legacy Imaging Survey DR10 and NUV, FUV from GALEX via VizieR. Where SDSS is available, we use the default $ugriz$ `model Mags'. If these magnitudes agree with the overlapping $griz$ bands in PANSTARRS, the PanSTARRS $y$ aperture magnitude is used. In cases where SDSS data is unavailable, the code queries for both PanSTARRS and DECAM optical data. All photometric data have been converted to AB magnitudes.

When collecting NIR data, the code attempts to pull two aperture options: `total' extrapolated ($m_{ext}$) and isophotal ($m_{iso}$) from the 2MASS extended source catalogue. Both the `total' and isophotal apertures measure the total flux of each host galaxy - they are large apertures, capturing lower surface brightness galaxy flux. The `total' aperture radius is based on the radial light distribution, and extends to a `four disk scale' surface brightness. This is integrated with the lower radial limit given by $m_{iso}$ and the upper limit determined by the light profile to derive the `total' flux of the galaxy. Where $m_{ext}$ is not available, the standard aperture $m_{iso}$ is used. 

The magnitudes available from the ALLWISE catalogue \citep{wright} come in a set of apertures of defined angular size, making it difficult to select a single aperture that works for all galaxies. Instead, we use unWISE, queried via the Legacy Survey. unWISE includes forced photometry using a bespoke aperture for each galaxy, based on the effective galaxy radius measured from SDSS. This ensures that the total MIR flux is included for each galaxy. The forced photometry also enables deblending of neighbouring galaxies, which may overlap in the large default apertures. It is worth noting that blending was generally much less of an issue in the optical/NIR due to much smaller point spread functions (PSFs). Although the PSF in GALEX is more comparable to WISE, the low spatial density of UV sources means that blending is not usually an issue at these wavelengths.

\subsection{SED Fits}

For modelling the TDE host galaxy light, we use \textsc{Prospector} \citep{leja}. \textsc{Prospector} is a stellar population synthesis code that uses FSPS \citep{conroy} to create synthetic spectra for stellar populations and fits the constructed spectral energy distributions (SEDs) as a function of model specified free parameters. For this study, we invoke a model with the following 9 free parameters: stellar mass, metallicity, a six-component star formation history, and a dust parameter controlling the optical depth from general interstellar dust throughout the galaxy. This is similar to that of \textsc{Prospector}-$\alpha$ detailed by \cite{leja} \citep[as per][]{ramsden}. Using dynamic nested sampling with \textsc{dynesty}, \textsc{Prospector} derives the posterior probability distributions of each free parameter. 

Recently, \cite{hammerstein2022} also used \textsc{Prospector} to model the SEDs for a number of our host galaxy sample. However, only optical and UV data was considered and a parametric SFH was used. There are a number of different SFH priors that can be used within \textsc{Prospector} - here we use non-parametric. While they can be more computationally expensive, non-parametric SFHs are more flexible. They can describe the full diversity of SFH profiles and allow for a better understanding of the age of the system. Comparatively, parametric SFHs may struggle to model complex behaviours, such as bursts of star formation and sudden quenching, and can force the fit into the wrong form. This could impact the recovery of SFH profiles and derivation of stellar population age, which in turn could result in biases in derived stellar masses \citep{leja19}. Compared to \cite{ramsden}, we have updated the prior used for our non-parametric SFH to a Dirichlet prior. \cite{leja19} test a number of different non-parametric priors and find that while all of them recover the input SFHs reasonably well, the fits are strongly dependent on the prior. \textsc{Prospector} fits using the Dirichlet prior were found to be the most effective in recovering the widest range of input SFHs. Within this treatment, the fraction of mass formed is allowed to vary within fixed time bins.

The remaining free parameters used in this model are well described in \citep{leja}. Here, we focus on the stellar mass results, marginalising over all other parameters. A follow-up study (Ramsden et al, in prep) will consider the SFHs of TDE host galaxies in more detail. 

\subsection{Bulge Masses}
\label{bulge}

To derive TDE host galaxy bulge masses from the total stellar masses returned by Prospector, we estimate the decomposition of galaxy light into bulge and disk components by taking the ratio of PanSTARRS PSF and Kron fluxes, as in \cite{wevers19}. Using the g-band fluxes from each, the PSF fit is used to approximate the flux from the central component of the galaxy — effectively treating the centre as a point source and serving as a proxy for the bulge. The Kron aperture captures the total galaxy flux. Using the luminosity ratio of the two provides the bulge-to-total mass ratio, $(B/T)_g$:

\begin{equation}
    (B/T)_g \approx 10^{0.4(m_{Kron} - m_{PSF})}.
\end{equation}

If any unphysical (B/T)g ratio are derived, this indicates an issue with the available photometry and targets are removed from the sample.

\begin{figure}
\centering
\includegraphics[width=\linewidth]{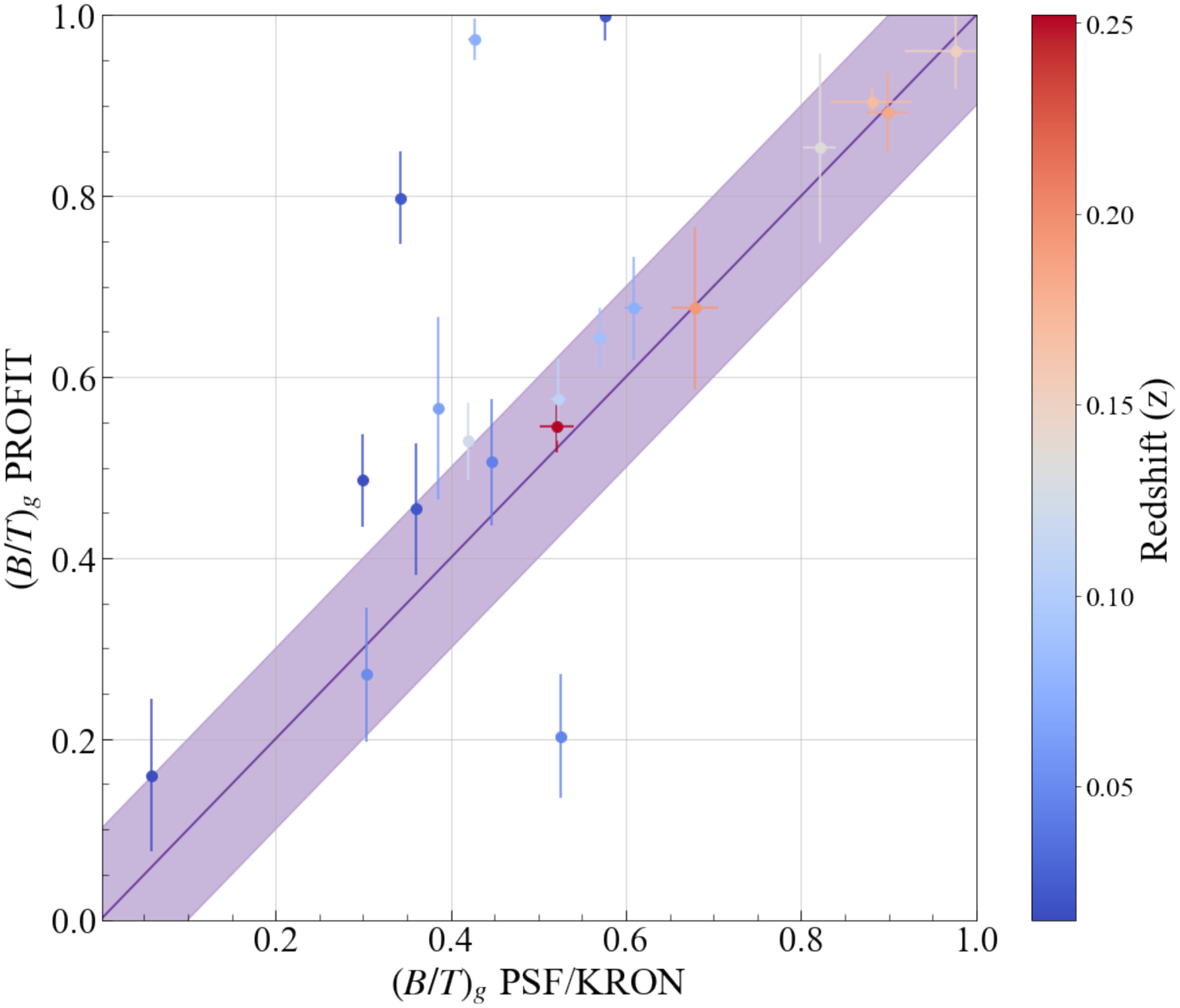}
\caption{A comparison of bulge-to-total mass ratio estimates derived using \textsc{profit} (Ramsden et al. (2022)) and those derived from PanSTARRS PSF/Kron fluxes, as used in this work. The $y=x$ fit is shown, with shaded regions indicating agreement within 10\%. Data points are coloured by redshift, with red representing higher redshift galaxies and blue representing lower.}
\label{bulge-comp}
\end{figure}

In our previous work, decomposition was done using \textsc{profit} \citep{robotham2017}, an $R$ package used for 2-D photometric galaxy profile modelling within a Bayesian framework \citep{ramsden}. With an ever increasing sample size, this method proves to be too computationally expensive. A future study will compare detailed star formation history profiles of a larger number of TDE host galaxies, with a matched sample of local galaxies for comparison. It would not be feasible to perform bulge-disk-decomposition with \textsc{Profit} on a sample of 700+ galaxies. Where available, we compare previous \textsc{profit} estimates to our PSF/Kron results and find that 70\% of our results have agreement within 10\%. Additionally, we find no dependence on target redshift in this comparison. This can be seen in Fig \ref{bulge-comp}.

The PSF/Kron ratio estimate assumes a centrally concentrated bulge with its light primarily captured by the PSF fit. In resolved galaxies, where the bulge is extended and not a perfect point source, this approximation may underestimate the bulge flux resulting in lower-limit estimates for the B/T ratio. This likely explains the marginal $\sim$ -0.1 dex offset observed in the PSF/Kron estimates relative to the PROFIT-derived values in Fig \ref{bulge-comp}. To assess the effect of this offset, we experimented with adding +0.1 dex to the PSF/Kron (B/T) ratio estimates before calculating bulge masses and analysing correlations with SMBH mass in \ref{data-analysis}. We found that the marginal shift had no significant impact on final relationships, and any changes in slope and offset remained within the uncertainties. 

\begin{figure*}
\centering
\includegraphics[width=1\textwidth]{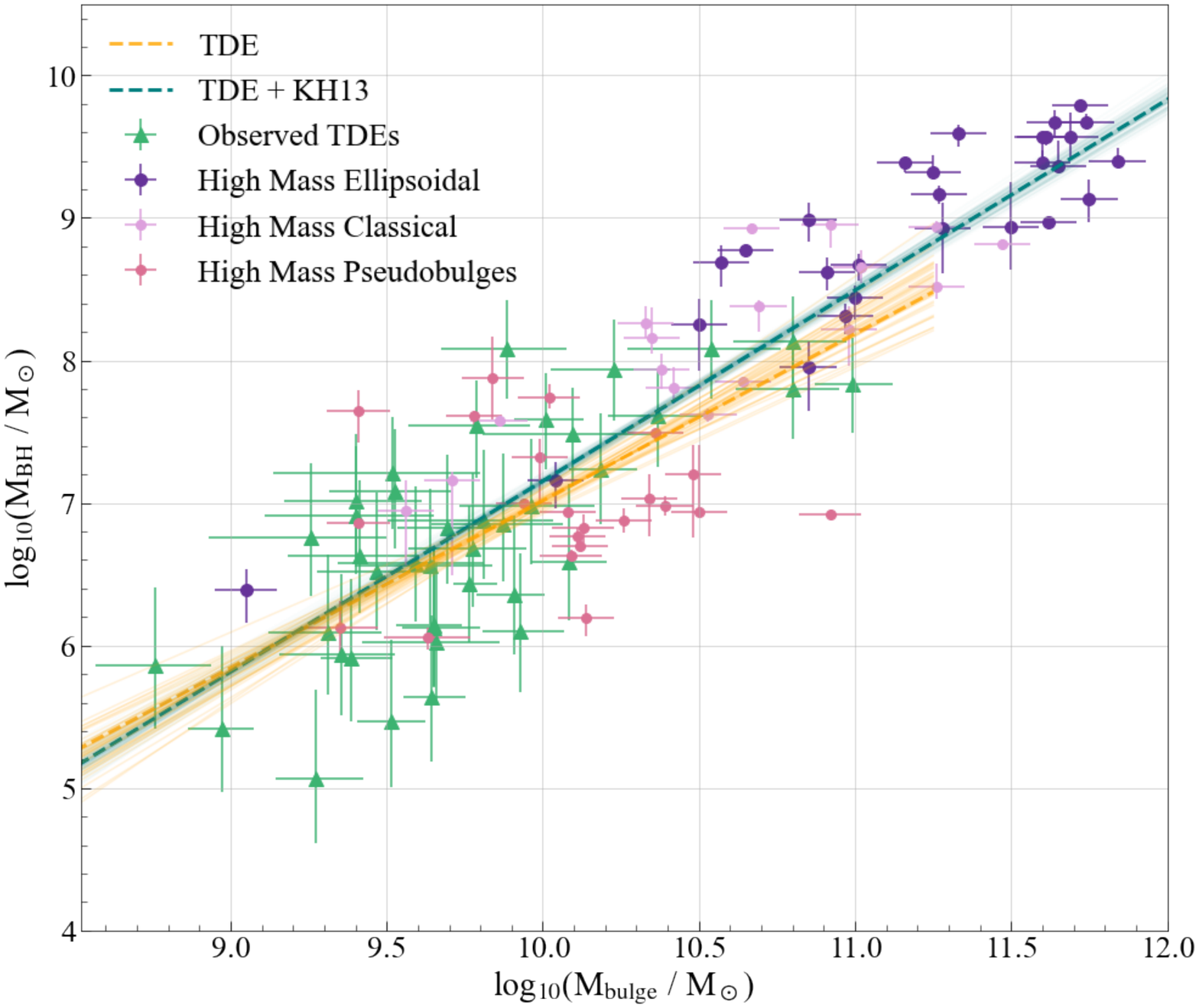}
\caption{SMBH mass as a function of host galaxy bulge mass for the TDE sample (green, showing statistical errors only) and the KH13 high-mass regime sample from Kormendy \& Ho (2013). This includes ellipsoidal galaxies (dark purple) and classical bulges (light purple), plus pseudobulges (pink), which were excluded from the original KH13 fit. The dashed teal line, Eq \ref{comb_fit}, shows the average fit to the TDE+KH13 sample after bootstrapping with perturbation, and the dashed orange line shows the equivalent for the TDE only sample, Eq \ref{tde_fit}. The shaded regions show the total range of bootstrapping with perturbation fits.}
\label{big_scatter}
\end{figure*}

\begin{table*}
\small
\begin{centering}
\caption{\small Correlation analysis for both the isolated TDE sample and combined sample.}
\label{t:stats}
\begin{tabular}{p{2cm} | p{3cm} | p{1.3cm}  | p{1.3cm} | p{1.3cm} | p{1.3cm} | p{1.3cm} | p{1.3cm} |}
\hline\hline
\multicolumn{2}{|c|}{} & $r_p$ & $P_p$ & $r_s$ & $P_s$ & m & c \\
\hline
\multirow{3}{=}{TDE only} & {Bootstrapping} & $0.684^{+0.040}_{-0.043}$ & $< 0.001$ & $0.645^{+0.049}_{-0.051}$ & $< 0.001$ &  $1.170^{+0.102}_{-0.100}$ & $-4.680^{+0.975}_{-1.002}$ \\ 
& \vfill  {Bootstrapping+Perturbation} &  \vfill $0.683^{+0.040}_{-0.040}$ & \vfill $< 0.001$ & \vfill $0.650^{+0.049}_{-0.051}$ & \vfill $< 0.001$ & \vfill $1.170^{+0.099}_{-0.096}$ & \vfill $-4.682^{+0.953}_{-0.979}$ \\ 
\hline
\multirow{3}{=}{TDE+KH13} & {Bootstrapping} & $0.873^{+0.008}_{-0.008}$ & $\ll 0.001$ & $0.853^{+0.011}_{-0.010}$ & $\ll 0.001$ & $1.338^{+0.027}_{-0.027}$ & $-6.226^{+0.290}_{-0.290}$ \\
& \vfill  {Bootstrapping+Perturbation} &  \vfill $0.873^{+0.008}_{-0.008}$ & \vfill $\ll 0.001$ & \vfill $0.853^{+0.010}_{-0.010}$ & \vfill $\ll 0.001$ & \vfill $1.338^{+0.026}_{-0.026}$ & \vfill $-6.228^{+0.269}_{-0.270}$ \\ 
\hline\hline
\end{tabular}
\end{centering}
\end{table*}

\section{The BH--Bulge Mass Relation}
\label{bh-bulge}

\subsection{Data Analysis}
\label{data-analysis}

When investigating any correlations in the data, we consider the strength of both linear and monotonic relationships using the Pearson Correlation Coefficient $r_p$ and the Spearman Rank Correlation Coefficient $r_s$. In common with previous calibrations of SMBH--Bulge mass relationships \citep[e.g.][]{kormendy,mcconnell}, we fit for power-law relations, i.e.  $\log (M_{\rm BH}) = A \log (M_{\rm bulge}) + C$, using standard $\chi^2$ minimisation. Any intrinsic scatter is calculated in the same manner as in \cite{kormendy} and \cite{mcconnell}. We use the Nukers' estimate, in which our scatter term, $\epsilon_0$ is optimised to ensure $\chi^2 / (\text{degrees of freedom}) = 1$ \citep{tremaine}.

To account for statistical errors within our data sets, bootstrapping methods are used. At every iteration, each $M_{\rm BH}$ - $M_{\rm bulge}$ pair is resampled by drawing from a 2D Gaussian distribution using the $1\sigma$ widths of the \textsc{Prospector} posteriors (propagated through to host galaxy bulge mass) and the $1\sigma$ errors on the plateau-derived SMBH masses from \citep{mummery2023}. Each error distribution is explored in full over 50,000 iterations, and probability distributions are produced for the statistical coefficients $r_p$ and $r_s$ and their p-values, and the best-fit gradients and intercepts. To ensure no single point carries significant weight and drives any identified correlation, we also apply perturbation \citep{curran}. Within each iteration, before resampling takes place, $N$ data points are randomly selected with replacement, where $N$ is the number of targets within our sample. This approach captures any effect of a limited sample size. 

See table \ref{t:stats} for all statistical values calculated.

\subsection{TDEs and the high mass regime}\label{obs_res}

Fig \ref{big_scatter} illustrates that TDEs are successful in calibrating the low mass end of the SMBH--stellar mass relationship. In order to compare TDE SMBHs to the known SMBH distribution at high masses, the observed TDE sample is combined with literature SMBH mass measurements covering the regime at higher masses. The high mass SMBH data consists of 67 targets sourced from \cite{kormendy}, which catalogues host galaxies and their central SMBH masses using measurements of different dynamics, such as stellar, maser disk, ionized gas, or CO molecular gas disk. Here, we retain the categorisation of morphological types - elliptical galaxies, classical bulges and pseudobulges. We exclude any data that \cite{kormendy} omitted in fitting, and any sources they identified as ongoing mergers.

Applying the bootstrapping with perturbation described in section \ref{data-analysis}, for the TDE only sample we find a Spearman coefficient $r_s = 0.65 (\pm 0.05)$ with significance $P_s < 0.001 $. The resampled best-fit straight line, has a gradient of $1.17 ({\pm 0.10})$ and a y-intercept of $-4.68 (^{+0.95}_{-0.98})$, which we can express as a power-law relation:

\begin{equation}\label{tde_fit}
\frac{M_{\text{BH}}}{10^9M_{\odot}} = (0.15^{+0.03}_{-0.03}) \cdot \left (\frac{M_{\rm bulge}}{10^{11}M_{\odot}}\right) ^{1.17\pm 0.10}; \qquad \epsilon_0 = 0.24 \text{ dex}. 
\end{equation}

Comparatively, the power-law relation for the combined TDE+KH13 sample has a gradient of $1.34(\pm 0.03)$ and a y-intercept of $-6.23 (^{+0.27}_{-0.27})$. Given the known correlation in the high-mass sample, it is unsurprising that we recover a strong correlation, with a Spearman coefficient of $r_s = 0.85 (\pm 0.01)$, and significance $P_s \ll 0.001$. This gives a SMBH--Bulge mass relationship: 

\begin{equation}\label{comb_fit}
\frac{M_{\text{BH}}}{10^9M_{\odot}} = (0.31^{+0.01}_{-0.01}) \cdot \left (\frac{M_{\rm bulge}}{10^{11}M_{\odot}}\right) ^{1.34\pm 0.03}; \qquad \epsilon_0 = 0.45 \text{ dex}. 
\end{equation}

Best-fits derived using bootstrapping with perturbation showed no significant difference to those derived with only resampling, as seen in Table \ref{t:stats}. This indicates that neither the TDE only or TDE+KH13 relationship was weighted by any extreme values in the data. All results within this paper are described using the holistic bootstrapping with perturbation method.

From Fig \ref{big_scatter}, it is clear that TDEs are successful in probing the low end of the mass spectrum and provide a reliable means for sourcing low mass SMBHs that may otherwise remain dormant. In previous work, \cite{ramsden} found a weak correlation between TDE SMBH masses derived using a \textsc{Mosfit} model of early-time optical/UV emission and host galaxy bulge masses obtained from a \textsc{Prospector} model similar to this work. In comparison, we now see a much more significant correlation. Assuming that there should be a strong relationship based on previous work in the high mass regime \citep{kormendy, mcconnell}, the SMBH masses derived with a late-time plateau emission model seem to more reliably recover the expected significant result.

\begin{figure}
    \centering
    \includegraphics[width=1\linewidth]{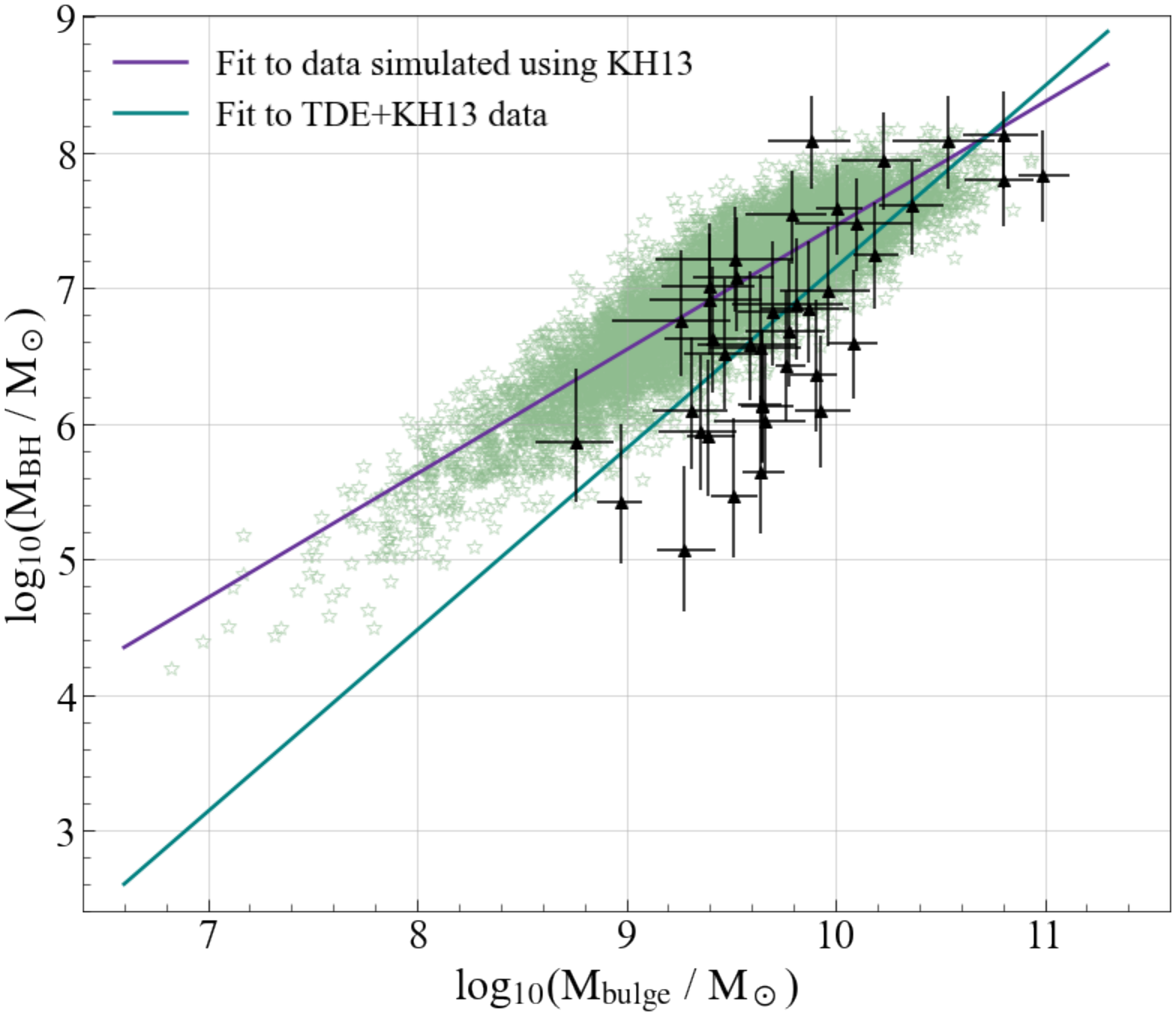} \\
    \includegraphics[width=1\linewidth]{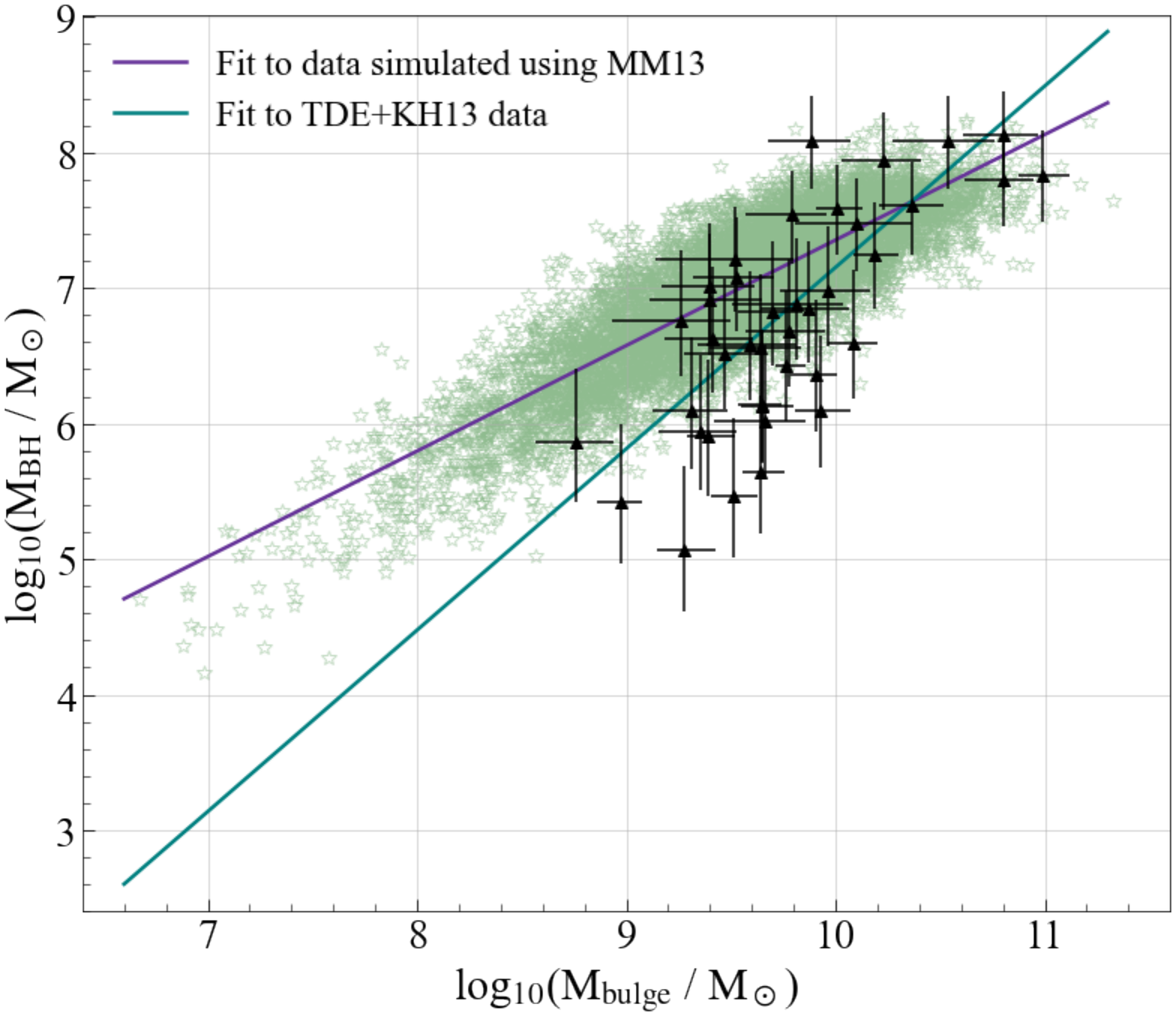}
    \caption{The simulated, observable TDE population produced when using the literature high-mass regime relationships. The simulated populations consider non-spinning SMBHs with an optical survey cut of 19.5 mag as described in section \ref{sim_method}. On the left, the simulation derives initial SMBH masses with the Kormendy \& Ho (2013) relationship and on the right, the Mcconnell \& Ma SMBH--Bulge relationship. The derived TDE+KH13 scaling relationship (teal) is compared with the linear fit to the simulated population (purple). The observed TDEs detailed in table \ref{obs_tdes} are overlayed.}
    \label{mm_sim}
\end{figure}

\begin{figure*}
\includegraphics[width=1.01\linewidth]{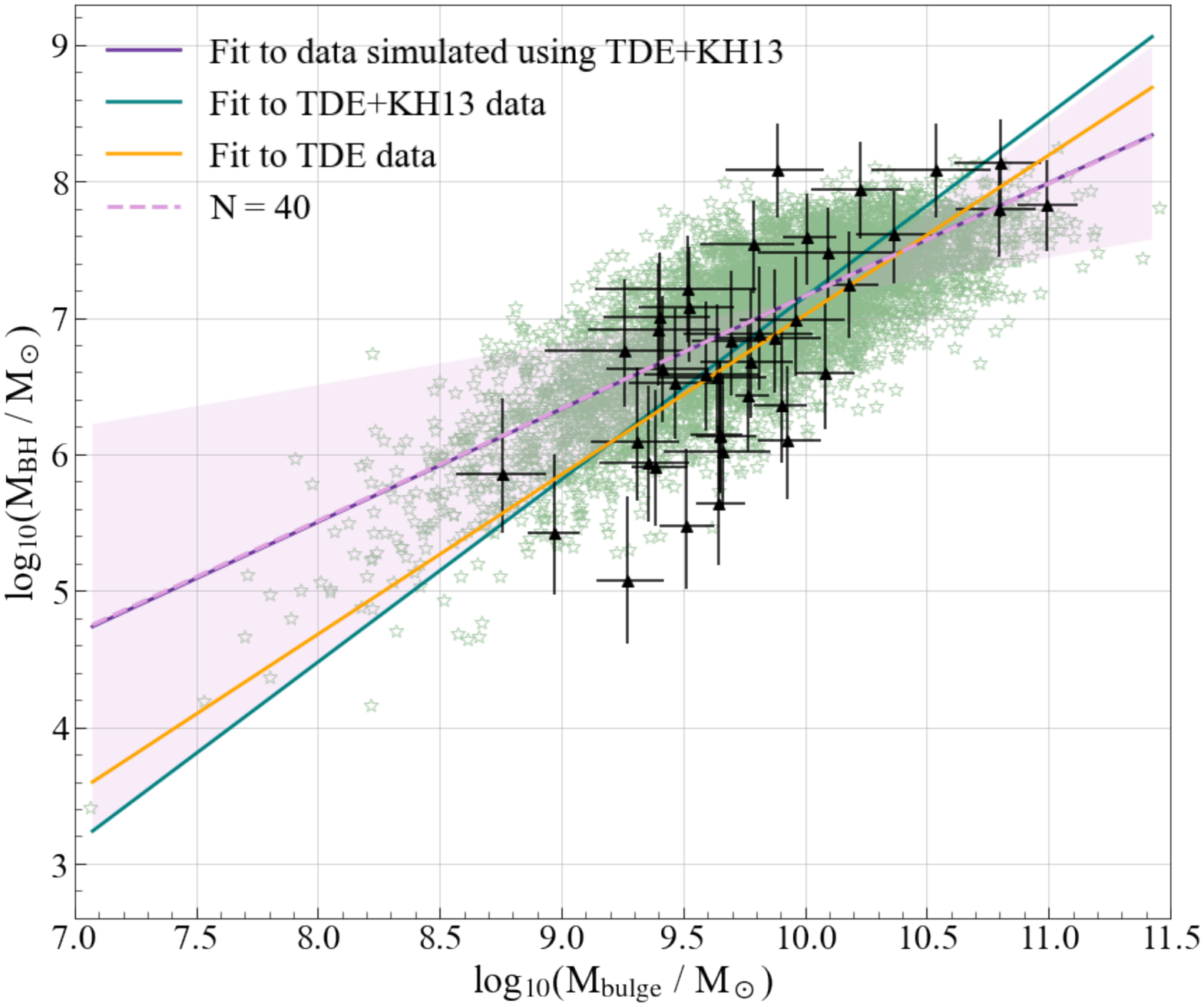}
\caption{The simulated, observable TDE population produced when using the TDE+KH13 SMBH--Bulge relationship (Eq \ref{comb_fit}). This simulated population considers non-spinning SMBHs and an optical survey cut of 19.5 mag as described in section \ref{sim_method}. The TDE+KH13 relationship (teal), is compared with the linear fit to the simulated population (purple). The derived  TDE only fit is shown in orange. For a more equitable comparison, the median and total range of fits to N=40 randomly sampled simulated TDEs, over 10,000 iterations, is shown (light purple and shaded region). The observed TDEs detailed in table \ref{obs_tdes} are overlayed.}
\label{comb_sim}
\end{figure*}

The newly derived SMBH--Bulge mass relationship, spanning a SMBH mass range $\sim 10^5 - 10^{10} M_{\odot}$ can be compared to both \cite{kormendy} and \cite{mcconnell}, see Table \ref{powerlaw-fits} for values of the coefficient, $\alpha$, and power-index, $\beta$.

\begin{table}
\small
\caption{\small Power-law fits to SMBH--Bulge mass from this work (TDE+KH13), Kormendy \& Ho 2013 (KH13) and McConnell \& Ma 2013 (MM13).}
\label{powerlaw-fits}
        \begin{center}
\renewcommand{\arraystretch}{1.6}
\begin{tabular}{l l l l} 
 \hline\hline
  & $\alpha$ & $\beta$ & $\epsilon_0$  \\ 
 \hline
 TDE+KH13 & $0.310^{+0.010}_{-0.010}$ & 
 $1.340^{+0.030}_{-0.030}$ & $0.45$ \\
 KH13 & $0.490^{+0.060}_{-0.050}$ & $1.160^{+0.080}_{-0.080}$ & $0.29$ \\
 MM13 & $0.290^{+0.003}_{-0.003}$ & $1.050^{+0.110}_{-0.110}$ & $0.34$ \\
  \hline\hline
\end{tabular}
\hypersetup{citecolor=blue}
\end{center}
\small \textbf{Note:} All SMBH--Bulge mass relations are of the form, \\ ($M_{\rm BH}/10^{9}M_{\odot}) = \alpha (M_{\rm bulge}/10^{11} M_{\odot})^{\beta}$ and $\epsilon_0$ is the intrinsic scatter.
\end{table}

Considering the values of the exponents as well as their uncertainties, we find some evidence for a steeper relation across the full TDE+KH13 population, compared to literature calibrations in the high-mass regime, at the level of $\gtrsim 2\sigma$. It appears that if a single slope extends for the whole BH mass range, it must be steeper to include the TDE population. The TDEs appear to be highly constraining at the lower end of the mass spectrum, anchoring the relationship. At the same time, the TDE population appears to smoothly join with the high-mass sample. 

Moreover, the so-called `pseudobulges', previously excluded from their final fit by \citet{kormendy}, overlap significantly with the TDE population. Pseudobulges contain central structures resembling those seen in classical bulge galaxies. However, observations reveal properties that are more disk-like such as flatter morphologies and a build-up of centrally concentrated components that mimic merger-built bulges. It has been suggested that pseudobulges are formed slowly through internal evolution of gas within the galactic disk \citep{kormendy, kormendy_n_kennicutt, fisher}.

When fitted with a Sérsic profile, pseudobulges indicate smaller bulge components ($n_b \lesssim 2$) and therefore lower mass SMBHs \citep{fisher}. Pseudobulges are known to have a tight SMBH--velocity dispersion relation \citep{hu}, though the slope is significantly shallower than that measured in classical bulges and elliptical galaxies. This suggests that their central SMBH growth is slower than that in early-type bulges \citep{hu, kormendy, jiang}. However, when combining the psuedobulges with the early-time bulges, the recovered fit is steeper than that of the early-time alone - the pseudobulges act as an anchor in the same way as the TDEs. Here, the combination of both TDEs and pseudobulges supports the need for a steeper relation if it is expected to extend down to lower masses. 

The above analysis makes a number of assumptions. In particular, we assume that stellar masses derived for the \cite{kormendy} galaxies are directly comparable to those derived with Prospector for the TDE host galaxies. In reality, choice of star formation history parameters and stellar population assumptions can significantly impact stellar mass estimates \citep{schutte,reines15}. To test the robustness of these results and assess the impact of our use of bulge masses, we repeat the above analysis for a total stellar mass sample. For this comparison, we re-compute total stellar masses for the KH13 high mass galaxies using Prospector, ensuring consistency in mass estimation across samples. This analysis is presented in Appendix \ref{a:A}, where we find that the key results remain unchanged.

\section{Forward Modelling of TDEs}
\label{simulation}

The scaling of SMBH mass with host galaxy bulge mass in the TDE-only sample remains statistically strong, but is shallower in comparison to that found in the TDE+KH13 sample. This could be due to a finite sample size, or could be the result of observational selection effects, such as the Hills and Malmquist biases. Here we use the term Hills bias to describe the impact of the critical Hills mass which biases TDEs towards SMBHs of mass $\lesssim 10^8 {\rm M}_{\odot}$. While this Hills bias is key to the success of TDEs in detecting low mass SMBHs, the event horizon suppression of the TDE rate due to the Hills bias may result in a flattening of the relation at high masses \citep{ramsden}. However, we note that maximally rotating SMBHs can soften this cut-off, in some cases increasing the maximum Hills mass to $10^9 M_{\odot}$ \citep{mummery2023b, kesden}. In addition, the Malmquist bias describes the preferential detection of intrinsically bright objects which appear to be more common as they can be observed over a greater volume. \cite{mummery2023} found that both the peak and plateau luminosities of TDEs scale positively with SMBH mass. This suggests that at greater distances, the lowest mass SMBHs may be missed depending on survey limits. This bias may cause an apparent flattening at the low end of the mass spectrum. It is therefore possible that the combination of Hills bias suppression at the high end of the TDE mass distribution and the Malmquist bias at the low end could cause the comparative flatness we see in the isolated TDE sample. 

\subsection{Simulation Method}
\label{sim_method}

Forward modelling of TDEs is used to determine whether the flatness in SMBH--Bulge mass scaling in the TDE only sample, compared to that in the TDE+KH13 sample, can be caused by these selection effects. We simulate a population of TDEs based on a number of random draws from priors on host galaxy mass, mass of the disrupted star and distance to the event. Here, sampling from the host galaxy mass, rather than directly from the BH mass, allows us to test the effect of different scaling relations. We use the sampled parameters to determine whether a given encounter produces a detectable TDE, given a survey limiting magnitude. This process accounts for the effects of both the Hills and Malmquist bias, as we describe below.

The total stellar mass of a galaxy is sampled from the Schechter mass function, as described by \cite{schechter},

\begin{equation}
    \Phi(M) = \text{ln}(10) \cdot \Phi^* \cdot [10^{(M-M^*)(1+\alpha)}] \cdot e^{[-10^{(M-M^*)}]}
\end{equation}

 where $\Phi^*$ is the normalisation, $M = \text{log}(M_{\text{stellar}}/M_{\odot})$ is the galaxy stellar mass, $M^*= \text{log}(M^*_{\text{stellar}}/M_{_\odot})$ is the mass at the break between the power-law and exponential parts of the distribution, and $\alpha$ is the slope at low-masses. Here we use a calibration based on all galaxy types over a redshift range of $0.2 \leq z < 0.5$, with $\Phi^* = 12.16 \cdot 10^{-4} \text{Mpc}^{-3}$, $M^* = 11.22$, and $\alpha = -1.29$, as in \cite{muzzin}. The sampled total stellar mass is converted into bulge mass using tabulated estimates of bulge to total mass $(B/T)_g$ ratios. Using a large sample of SDSS galaxies \citep{vanvelzen2018}, the average $(B/T)_g$ ratio for nine bins of total stellar mass is shown in table B1 in \cite{stone18}. Here we use linear interpolation between stellar mass bins, to draw the $(B/T)_g$ estimate for the sampled total stellar mass, and derive the host galaxy bulge mass. Using the bulge mass, we then draw SMBH mass from one of the SMBH--Bulge scaling relationships quoted in Table \ref{powerlaw-fits}: TDE+KH13, Eq \ref{comb_fit}; \cite{kormendy} or \cite{mcconnell}, with stated intrinsic scatter to reflect any uncertainty in results.
 
This method assumes that the $(B/T)_g$ distribution of the general population is reflective of that for TDE hosts. In reality, TDE hosts may show a preference for bulge-dominated galaxies. However, because we will sample BH masses based on bulge scaling relations, a systematically higher B/T ratio for our simulated population would have no effect on the recovered slope, but will simply weight the bulk of the population towards higher bulge \emph{and} BH mass.

Separately, we consider the mass of the disrupted star. The stellar mass is sampled from the Kroupa initial mass function (IMF), which is a broken power-law:

\begin{equation}
\begin{split}
    & \xi(m) \propto m^{-\alpha_i} \\[10pt]
    & \alpha_0 = 2.3, \qquad 0.50 \leq m / M_{\odot}, \\
    & \alpha_1 = 1.3, \qquad 0.08 \leq m / M_{\odot} < 0.50, \\
    & \alpha_2 = 0.3, \qquad 0.01 \leq m / M_{\odot} < 0.08
\end{split}
\end{equation}

\vspace{2mm}

The limiting Hills mass - the maximum SMBH mass around which a star of given stellar mass can be disrupted outside the event horizon - is calculated for each star using the derivation from \cite{mummery2023}. We assume that the stars approach in the equatorial plane of the black hole, a simplification which neglects the effects of orbital inclination, but which  includes the effect of SMBH spin:

\begin{equation}
    M_{\bullet} = \left( \frac{5c^6R^3_*}{G^3M_*} \right)^{1/2} \cdot \frac{1}{(1 + \sqrt{1-a_{\bullet}})^3}
\end{equation}

Where $R_*$ and $M_*$ are the disrupted stellar radius and mass respectively, and $a_{\bullet}$ is SMBH spin, which we set to $a_{\bullet} = 0$ or use a flat prior where $-1 \leq a \leq 1$. Note that the stellar radius is calculated using the mass-radius relationship of \cite{kippen}: 

\begin{equation}
 R \propto
 \begin{cases}
      R_{\odot} (M_* / M_{\odot})^{0.56}, \qquad M_* \leq M_{\odot}\\

      \\
      
      R_{\odot} (M_* / M_{\odot})^{0.79}, \qquad M_* > M_{\odot}\\
    \end{cases} 
\end{equation}

If the calculated Hills mass is larger than the sampled SMBH mass, theoretically the TDE can be observed, and we next consider the Malmquist luminosity bias. Otherwise, if the Hills mass is smaller than the sampled SMBH mass, the event would not be observable and so it is not considered further. 

\cite{mummery2023} discovered a significant correlation between the g-band peak luminosity of a TDE and its black hole mass. Sampling within the uncertainties on this relation allows us to estimate the peak luminosity for each TDE that passed the Hills bias cuts: 

\begin{equation}
    \text{log}_{10}(\nu L_{\nu,\text{peak}}) = \frac{ \text{log}_{10}(M_{\rm BH}/M_{\odot}) - 6.52 }{0.98} - 43 \qquad \epsilon_0 = 0.53 \text{ dex}
\end{equation}

In order to convert this peak luminosity to a $g$-band apparent magnitude, a distance D is randomly sampled from a probability distribution of $P(D) \propto D^2$, over $0 < D < 1000$ Mpc. An optical survey limit of 19.5 mag is then set to filter for the final sample of observable TDEs according to current survey limits. From \cite{yao23}, it can be assumed that ZTF is spectroscopically complete for TDE candidates brighter that $\sim 19$ mag, though several TDEs detailed in this sample and that in \cite{hammerstein2022} have mag $\sim 19.5$. Only TDEs brighter than the survey limiting magnitude were retained. 

It is worth highlighting that the forward modelling simulation applies a cut based on TDE peak luminosity as opposed to plateau luminosity. This approach aims to mimic the detection strategy employed by ZTF, which discovers TDEs based on their early-time optical/UV peak emission in real time. The late-time plateau becomes relevant post detection during follow-up. \cite{mummery2023} find that the plateau luminosity is typically $\sim 1\%$ of the peak luminosity. Therefore for TDEs peaking brighter than 19.5 mag, the late-time emission should be brighter than 24.5 mag. While these plateaus will almost always be too faint to recover in the ZTF single-epoch flux limit, it has been shown that stacking $\sim 100$ ZTF observations makes it possible to recover plateaus at $\sim 24$ mag \citep{mummery2023}.

\begin{figure}
\includegraphics[width=0.49\textwidth]{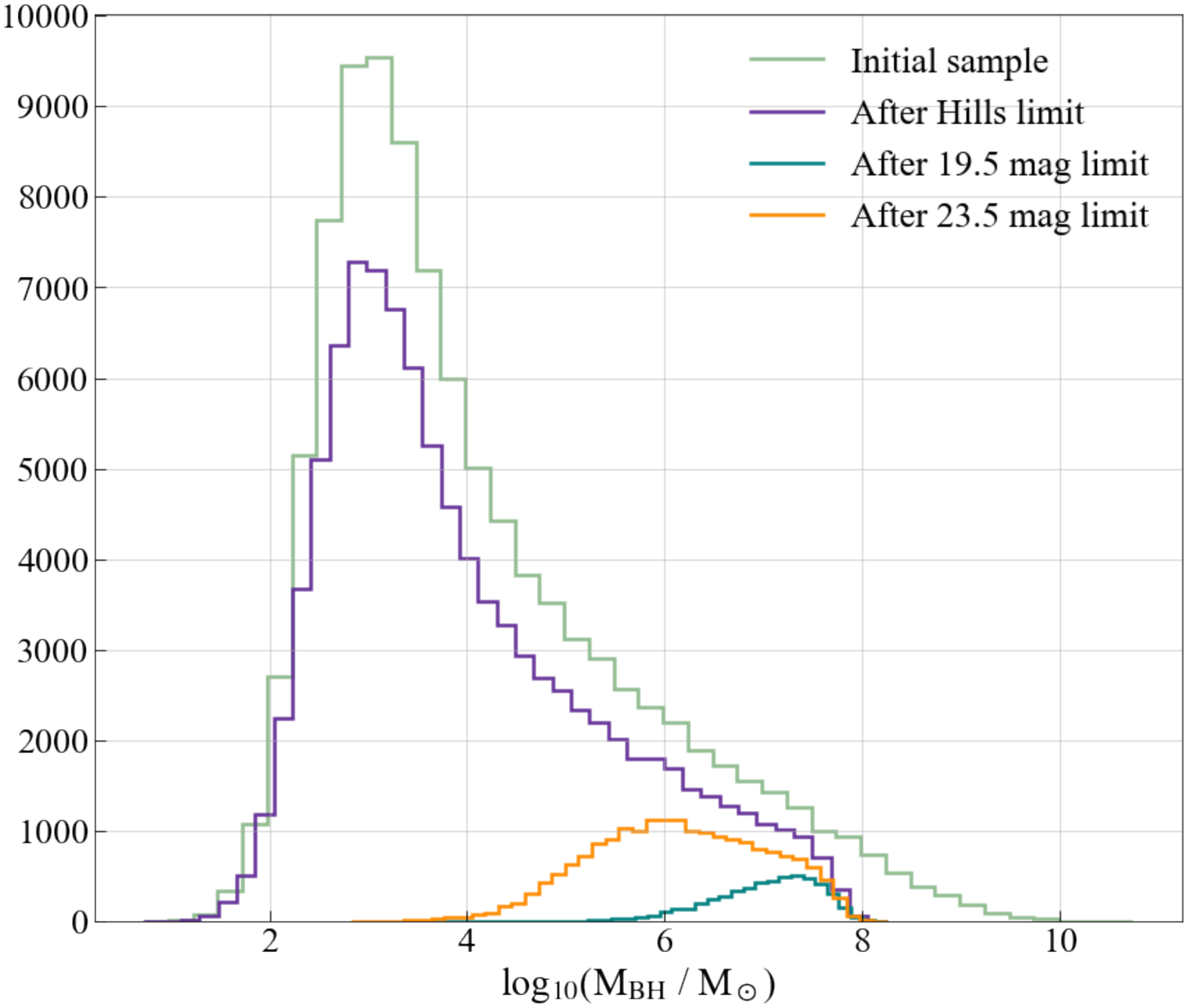}
\caption{Distribution of simulated non-spinning SMBH masses after initial draws (green), Hills mass cuts (purple) and Malmquist bias cuts for current survey limits of 19.5 mag (blue) and LSST survey limits of 23.5 mag (orange).}
\label{distributions}
\end{figure}

\begin{figure*}
    \centering
    \includegraphics[width=0.49\linewidth]{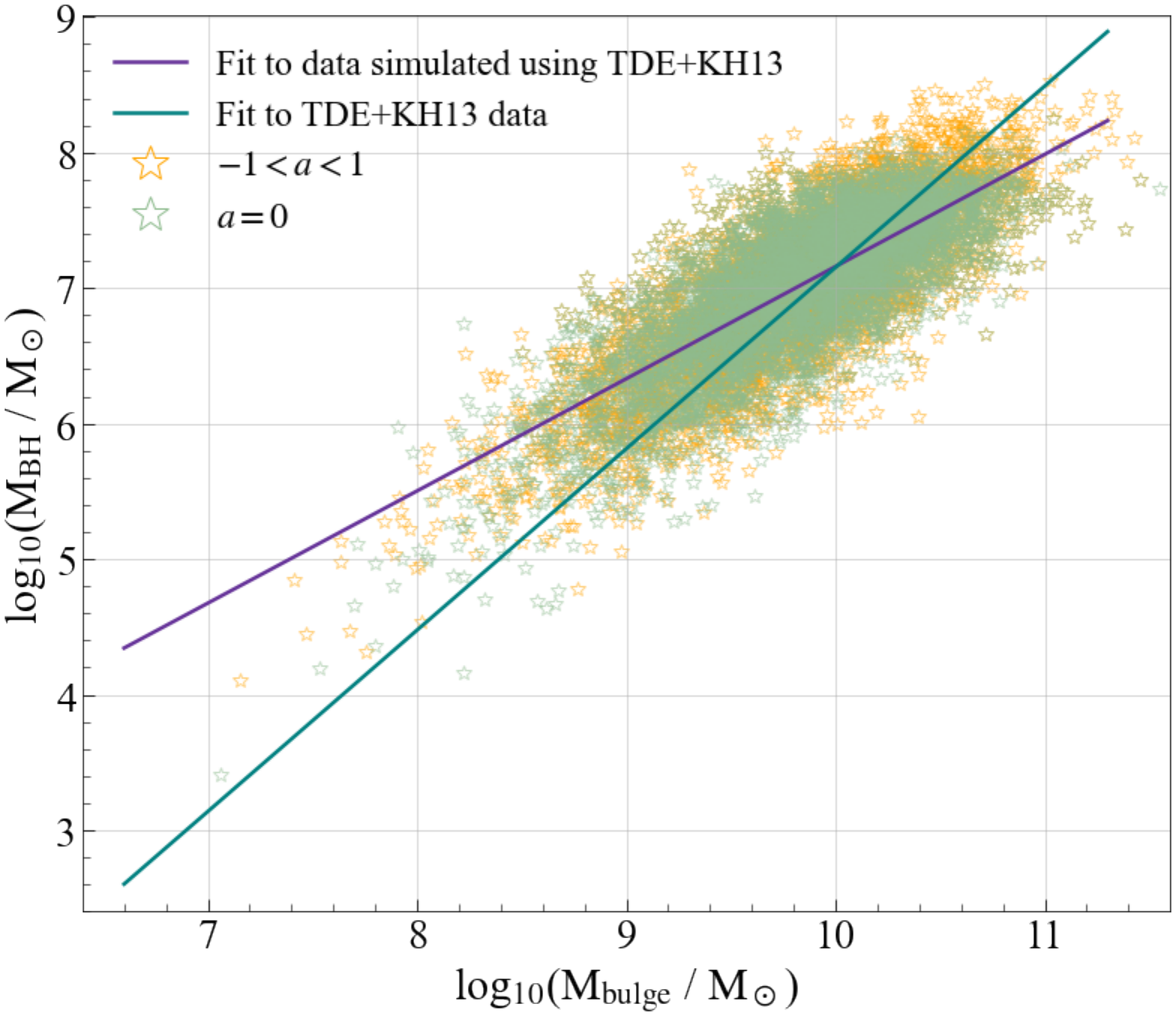}
    \includegraphics[width=0.49\linewidth]{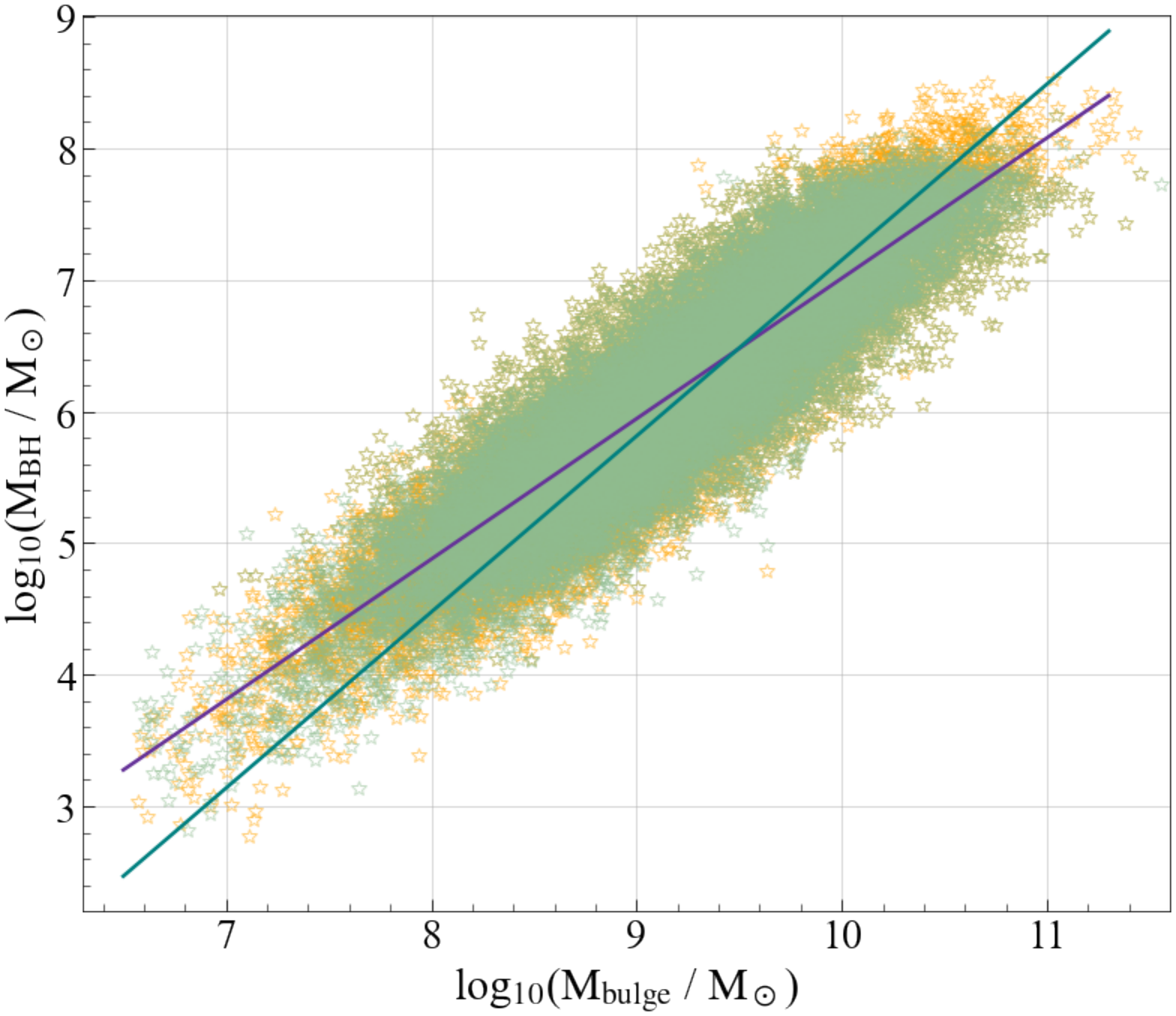}
    \caption{The simulated, observable TDE population for both non-spinning (green) and spinning (orange) SMBHs as described in section \ref{simulation}, considering different survey magnitude limits. Two simulated samples are shown, on the left a current magnitude cut of 19.5 mag and on the right a future cut of 23.5 mag. The solid purple lines are the best-fits to the non-spinning simulated samples. Note that there is no significant difference for the spinning simulated sample. The solid teal line shows the expected TDE+KH13 relationship, Eq \ref{comb_fit}.}
    \label{19.5v23.5}
\end{figure*}

\subsection{Application to Correlation}

In section \ref{obs_res}, we found that relative to previous calibrations of BH scaling relations (KH13, \cite{kormendy}; MM13, \cite{mcconnell}) our TDE-only population showed a flatter relation, while the full population showed an apparently steeper relation. In principle, both Hills and Malmquist bias could help to account for the former, while a strong Hills suppression of high-mass TDEs could potentially account for the latter. To test whether these selection effects are responsible for our results, we first simulate a population of observable TDEs by assuming that high-mass SMBH--Bulge mass scaling relations \citep[e.g.][]{kormendy, mcconnell}, apply over the full mass range. The results are shown in Fig \ref{mm_sim}. While these simulations do show a flattening of the TDE only relation as observed, they do not reproduce the observed TDE population. The observed TDEs sit systematically below the simulated population even when the physical and observational biases are considered. In other words, the effect of event horizon suppression is not significant across the full TDE bulge mass range, making a difference only at the very high end, and so cannot account for the steeper relationship we find in fitting the full TDE+KH13 data set.

The above finding suggests that an intrinsically steeper scaling relation is needed to account for the BH and bulge mass distributions of TDEs. Therefore, we consider the steeper combined TDE+KH13 relationship derived in Eq \ref{comb_fit}, and use this for forward modelling. We note that this relationship includes the observed TDE sample that has already been impacted by both the Hills and Malmquist biases. However, as shown above, these selection effects predominantly impact the extremes of any TDE sample. The Hills bias reduces the number of high mass SMBHs observed, due to the critical Hills mass limit, and the Malmquist removes low luminosity and therefore low mass SMBHs, due to restrictions on survey resolution. The combination of these effects influence the gradient of the SMBH--Bulge mass relationship \emph{within} the TDE population, but have little effect on where the bulk of the sample (TDEs with BH mass in the $10^6 < M_{\rm BH} < 10^8$ range) sits in the parameter space. 

It can be seen in Fig \ref{comb_sim} that when we use the TDE+KH13 relationship in our simulation, the observed TDE sample is successfully recovered and the relative flatness in the TDE-only sample is accounted for. If we fit a scaling relation to this simulated population of $\gtrsim5000$ TDEs above a limiting magnitude of 19.5, we find:

\begin{equation}\label{tde_sim_fit19.5}
\frac{M_{\text{BH}}}{10^9M_{\odot}} = (0.10) \cdot \left (\frac{M_{\rm bulge}}{10^{11}M_{\odot}}\right) ^{0.83}    
\end{equation}

which agrees with the observed flattening seen in Eq \ref{tde_fit} to within $\sim 3\sigma$. For a more equitable comparison, we calculate the best-fit straight line to a simulated population of only 40 TDEs (matching the size of the observed TDE data set) and repeat this 10,000 times. The range of possible fits is illustrated as the shaded region in Fig \ref{comb_sim}, where the the power-law index ranges within $0.31 < \beta < 1.32$, encompassing our TDE only fit. It is clear that regardless of which 40 TDEs are selected, the resulting relationship is flatter than when fitting to the full TDE+KH13 data set. Thus, the comparative flatness appears to be consistent with the combination of the Hills and the Malmquist biases.

Provided we are not missing any other selection effects, and that the full spectrum of SMBH masses lie on a single power-law scaling relation, our results provide preliminary evidence for a steeper scaling than has previously been suggested. Alternately, it is possible that the a break in the power law could exist and the relation may change to a steeper slope at low mass. If a break in the scaling at low mass did exist, it could provide important constrains on SMBH growth channels \citep{greene20}. However, we reiterate that our steeper relation is significant only at the $\gtrsim 2\sigma$ level with the current TDE sample, and without a larger TDE sample we cannot confidently confirm either scenario.

\section{Future Outlook}
\label{future}

With a wide and dense coverage, the forthcoming optical survey LSST is predicted to increase the TDE detection rate to $> 1000 \text{ year}^{-1}$ \citep{lsst,bricman}. Uniformly observing an 18,000 deg$^2$ region, the survey will produce adjacent and overlapping imaging of over half the sky. In a single visit, the expected 5$\sigma$ depth in the g-band is 25.0 mag - this is set to increase to 27.4 mag in the final co-added image stack \citep{bianco}. To determine the impact of LSST on TDE scaling relations, we simulate a population of TDEs with a magnitude limit of 23.5, implicitly assuming that any sources peaking $>1$ mag above the survey limit can be confidently identified as TDEs. Fig \ref{distributions} shows the effect of Hills and Malmquist bias on the recovered BH mass distributions for the same simulated population of TDEs but using different survey limits of 19.5 vs 23.5 mag. At the high-mass end, Hills suppression introduces the sharp cut-off similarly for both populations as expected. At the low end, the difference is stark. Malmquist bias kicks in quickly for the shallow survey, whereas the deep survey traces the shape of the input population down to $\approx 10^6$ $M_\odot$. The 90\% ranges of the distributions are $10^{5.99 - 7.74} M_{\odot}$ for the 19.5 mag survey and $10^{4.71 - 7.53} M_{\odot}$ for the 23.5 mag survey.

The depth of LSST will also help in recovering TDE plateaus. At the distance limit of our simulation, $z=0.2$, the final stacked depth of 27.4\,mag in $g$ corresponds to $\nu L_\nu\approx10^{40}$\,erg\,s$^{-1}$. This would enable plateaus to be detected for many TDEs with BH masses down to $\sim10^{4}$\,M$_\odot$. In the case of a non-detection of a plateau with LSST, JWST follow-up may be an efficient way to find intermediate-mass BHs down to $\sim 10^{3}$\,M$_\odot$. Compared to what exists now, this will be extremely powerful in improving scaling relation calibrations and extending to very low masses. This is necessary in order to confirm our finding of a steeper scaling relation. With a well calibrated scaling relation using LSST TDEs, unprecedented survey depth will enable us to use the very deep LSST galaxy photometry to estimate BH masses for the thousands of TDEs at higher redshifts, for which plateau-based mass estimates are not always possible. This will be key for understanding TDE physics and rates as a function of BH mass.

We therefore investigate how much we can reduce the Malmquist bias in calibrating scaling relations when we detect TDEs down to LSST magnitude limits. Fitting a power-law to this simulated population gives:

\begin{equation}\label{tde_sim_fit23.5}
    \frac{M_{\rm BH}}{10^9 M_{\odot}} = (0.12) \cdot \left (\frac{M_{\rm bulge}}{10^{11}M_{\odot}} \right )^{1.07}
\end{equation}

As shown in Figure \ref{19.5v23.5}, this fit is much closer to the input slope used in our simulation, whereas the simulation with a survey limit of 19.5 mag produces an artificially shallow slope. The cut-off at high SMBH mass is softened when SMBH spin is considered, though this does not make a substantial difference to the observed slope of the scaling compared to the effect of the Malmquist bias. For a maximally rotating SMBH ($a = 1$), the Hills mass can increase by up to a factor of 8. If the inclination of the incoming star’s orbit is also considered, the Hills mass can increase even further, by up to a factor of $\sim 12$ \citep{mummery2023b}. The impact of this softening can be seen in the extension of the observable TDE samples to higher masses in Fig \ref{19.5v23.5}. 

Instead, by directly highlighting low mass BHs that would otherwise lay dormant, TDEs could be a powerful tool in determining the true occupation fraction. Understanding this fraction at lower masses and determining whether a steeper overall SMBH--Bulge mass relationship is required, or if a power-law break at lower masses exists, would further our understanding of SMBH evolutionary pathways. It is predicted that any differences we see in the occupation fraction or black hole-galaxy scaling relations, at lower masses compared to higher masses, may be an indication of dominate SMBH seeding channels. Current theories suggest that seed black holes may be the result of: the deaths of early-time first stars (Pop III stars); the direct collapse of gas clouds; or gravitational runaway events within globular clusters \citep{greene20}. By successfully probing low mass SMBHs, future samples of TDEs will be key to calibrating this known SMBH--Bulge mass scaling relationship further and improving our understanding of SMBH seeding mechanisms. 

\section{Conclusions}
\label{conclusion}

Using \textsc{Prospector} the photometric data of 40 plateau TDE host galaxies were modelled. The derived host galaxy bulge mass for each event was paired with SMBH masses derived in \cite{mummery2023} using a new late-time optical/UV emission model. The SMBH--Bulge mass relation for the observed TDEs, both alone and in combination with the high-mass regime from \cite{kormendy} was investigated. Forward modelling of TDEs was later used to consider the impact of physical and observational selection effects on our derived relationships.

Our main conclusions are as follows:

\begin{itemize}
    \item We confirm that SMBH masses derived from late-time optical/UV plateau emission  correlate well with host galaxy bulge mass, and find that tidal disruptions events are successful in extending the known SMBH--Bulge mass relation to lower mass.

    \item We find host galaxy bulge masses in the range of $10^{8.76 - 10.99} M_{\odot}$, with a median of $10^{9.77} M_{\odot}$.

    \item A significant correlation between TDE SMBH mass and host galaxy bulge mass was identified. After accounting for observational uncertainties and a finite sample size, we measure a power-law slope of $1.17 \pm 0.10$. 

    \item When combining the observed TDE sample with that of the high mass regime, we recover a strong positive correlation. TDEs anchor the relationship at the low mass-end, and we find a slope of $1.34 \pm 0.03$. 

    \item If a single power-law relation does describe BH-bulge scaling across this mass range, we find marginal ($2\sigma$) evidence for a steeper relation than previous calibrations when including TDEs.

    \item As in previous work, the TDE-only relation is comparatively flatter than the relation including the full TDE+KH13 data set. Forward modelling is used to consider the impact of selection effects, such as Hills and Malmquist biases, and finds that the shallower slope within the TDE population can be explained by these effects.
\end{itemize}

TDEs have again proved to be promising probes for low mass SMBHs. Future detection rates are set to increase tenfold with the upcoming LSST survey, and improved understanding of the SMBH--Bulge mass relationship at the low end of the mass spectrum will soften the need for follow-up observations and allow us to estimate the black hole mass density function. When late-time Optical/UV plateau detections do become readily available and provide independently derived SMBH masses, TDEs will be key to further calibrating the relationship and aid in improving our understanding of black hole formation paths.

\section*{Acknowledgements}

We thank an anonymous referee for their comments that improved the manuscript.

PR acknowledges support from STFC grant 2742655.

MN is supported by the European Research Council (ERC) under the European Union’s Horizon 2020 research and innovation programme (grant agreement No.~948381) and by UK Space Agency Grant No.~ST/Y000692/1.

SM is supported by UK Space Agency Grants No. ST/Y005201/1 and ST/Y000692/1.

AM is supported by a Leverhulme Trust International Professorship grant [number LIP-202-014].

\section*{Data Availability}

This article is based on existing publicly available data.



\bibliographystyle{mnras}
\bibliography{refs} 

\begin{thebibliography}{}
\makeatletter
\relax
\def\mn@urlcharsother{\let\do\@makeother \do\$\do\&\do\#\do\^\do\_\do\%\do\~}
\def\mn@doi{\begingroup\mn@urlcharsother \@ifnextchar [ {\mn@doi@} {\mn@doi@[]}}
\def\mn@doi@[#1]#2{\def\@tempa{#1}\ifx\@tempa\@empty \href {http://dx.doi.org/#2} {doi:#2}\else \href {http://dx.doi.org/#2} {#1}\fi \endgroup}
\def\mn@eprint#1#2{\mn@eprint@#1:#2::\@nil}
\def\mn@eprint@arXiv#1{\href {http://arxiv.org/abs/#1} {{\tt arXiv:#1}}}
\def\mn@eprint@dblp#1{\href {http://dblp.uni-trier.de/rec/bibtex/#1.xml} {dblp:#1}}
\def\mn@eprint@#1:#2:#3:#4\@nil{\def\@tempa {#1}\def\@tempb {#2}\def\@tempc {#3}\ifx \@tempc \@empty \let \@tempc \@tempb \let \@tempb \@tempa \fi \ifx \@tempb \@empty \def\@tempb {arXiv}\fi \@ifundefined {mn@eprint@\@tempb}{\@tempb:\@tempc}{\expandafter \expandafter \csname mn@eprint@\@tempb\endcsname \expandafter{\@tempc}}}

\bibitem[\protect\citeauthoryear{{Bentz} et~al.,}{{Bentz} et~al.}{2013}]{Bentz2013}
{Bentz} M.~C.,  et~al., 2013, \mn@doi [\apj] {10.1088/0004-637X/767/2/149}, \href {https://ui.adsabs.harvard.edu/abs/2013ApJ...767..149B} {767, 149}

\bibitem[\protect\citeauthoryear{Bianco et~al.,}{Bianco et~al.}{2021}]{bianco}
Bianco F.~B.,  et~al., 2021, \mn@doi [The Astrophysical Journal Supplement Series] {10.3847/1538-4365/ac3e72}, 258, 1

\bibitem[\protect\citeauthoryear{Bricman \& Gomboc}{Bricman \& Gomboc}{2020}]{bricman}
Bricman K.,  Gomboc A.,  2020, \mn@doi [The Astrophysical Journal] {10.3847/1538-4357/ab6989}, 890, 73

\bibitem[\protect\citeauthoryear{{Chambers}}{{Chambers}}{2006}]{panstarrs}
{Chambers} K.,  2006, in {Ryan} S.,  ed., The Advanced Maui Optical and Space Surveillance Technologies Conference. p.~E39

\bibitem[\protect\citeauthoryear{Conroy, Gunn  \& White}{Conroy et~al.}{2009}]{conroy}
Conroy C.,  Gunn J.~E.,   White M.,  2009, \mn@doi [The Astrophysical Journal] {https://doi.org/10.1088/0004-637X/699/1/486}, 699, 486

\bibitem[\protect\citeauthoryear{Curran}{Curran}{2014}]{curran}
Curran P.~A.,  2014, arXiv: Instrumentation and Methods for Astrophysics

\bibitem[\protect\citeauthoryear{{Ferrarese} \& {Merritt}}{{Ferrarese} \& {Merritt}}{2000}]{ferrarese}
{Ferrarese} L.,  {Merritt} D.,  2000, \mn@doi [\apjl] {10.1086/312838}, \href {https://ui.adsabs.harvard.edu/abs/2000ApJ...539L...9F} {539, L9}

\bibitem[\protect\citeauthoryear{Fisher \& Drory}{Fisher \& Drory}{2008}]{fisher}
Fisher D.~B.,  Drory N.,  2008, \mn@doi [The Astronomical Journal] {https://doi.org/10.1088/0004-6256/136/2/773}, 136, 773

\bibitem[\protect\citeauthoryear{{Gebhardt} et~al.,}{{Gebhardt} et~al.}{2000}]{Gebhardt2000}
{Gebhardt} K.,  et~al., 2000, \mn@doi [\apjl] {10.1086/312840}, \href {https://ui.adsabs.harvard.edu/abs/2000ApJ...539L..13G} {539, L13}

\bibitem[\protect\citeauthoryear{{Greene} et~al.,}{{Greene} et~al.}{2016}]{greene16}
{Greene} J.~E.,  et~al., 2016, \mn@doi [\apjl] {10.3847/2041-8205/826/2/L32}, \href {https://ui.adsabs.harvard.edu/abs/2016ApJ...826L..32G} {826, L32}

\bibitem[\protect\citeauthoryear{{Greene}, {Strader}  \& {Ho}}{{Greene} et~al.}{2020}]{greene20}
{Greene} J.~E.,  {Strader} J.,   {Ho} L.~C.,  2020, \mn@doi [\araa] {10.1146/annurev-astro-032620-021835}, \href {https://ui.adsabs.harvard.edu/abs/2020ARA&A..58..257G} {58, 257}

\bibitem[\protect\citeauthoryear{{G{\"u}ltekin} et~al.,}{{G{\"u}ltekin} et~al.}{2009}]{Gultekin2009}
{G{\"u}ltekin} K.,  et~al., 2009, \mn@doi [\apj] {10.1088/0004-637X/698/1/198}, \href {https://ui.adsabs.harvard.edu/abs/2009ApJ...698..198G} {698, 198}

\bibitem[\protect\citeauthoryear{Hammerstein et~al.,}{Hammerstein et~al.}{2022}]{hammerstein2022}
Hammerstein E.,  et~al., 2022, \mn@doi [The Astrophysical Journal] {10.3847/1538-4357/aca283}, 942, 9

\bibitem[\protect\citeauthoryear{Hills}{Hills}{1975}]{hills}
Hills J.~G.,  1975, \mn@doi [Nature] {https://doi.org/10.1038/254295a0}, 254, 295–298

\bibitem[\protect\citeauthoryear{Honscheid \& DePoy}{Honscheid \& DePoy}{2008}]{decam}
Honscheid K.,  DePoy D.~L.,  2008, The Dark Energy Camera (DECam) (\mn@eprint {arXiv} {0810.3600}), \url {https://arxiv.org/abs/0810.3600}

\bibitem[\protect\citeauthoryear{Hu}{Hu}{2008}]{hu}
Hu J.,  2008, \mn@doi [Monthly Notices of the Royal Astronomical Society] {https://doi.org/10.1111/j.1365-2966.2008.13195.x}, 386, 2242–2252

\bibitem[\protect\citeauthoryear{Hung et~al.,}{Hung et~al.}{2017}]{hung}
Hung T.,  et~al., 2017, \mn@doi [The Astrophysical Journal] {https://doi.org/10.3847/1538-4357/aa7337}, 842, 29

\bibitem[\protect\citeauthoryear{Häring \& Rix}{Häring \& Rix}{2014}]{haring}
Häring N.,  Rix H.-W.,  2014, \mn@doi [The Astrophysical Journal] {10.1086/383567}, 487, L89–L92

\bibitem[\protect\citeauthoryear{{Ivezi{\'c}} et~al.,}{{Ivezi{\'c}} et~al.}{2019}]{lsst}
{Ivezi{\'c}} {\v{Z}}.,  et~al., 2019, \mn@doi [\apj] {10.3847/1538-4357/ab042c}, \href {https://ui.adsabs.harvard.edu/abs/2019ApJ...873..111I} {873, 111}

\bibitem[\protect\citeauthoryear{{Jiang}, {Greene}  \& {Ho}}{{Jiang} et~al.}{2011}]{jiang}
{Jiang} Y.-F.,  {Greene} J.~E.,   {Ho} L.~C.,  2011, \mn@doi [\apjl] {10.1088/2041-8205/737/2/L45}, \href {https://ui.adsabs.harvard.edu/abs/2011ApJ...737L..45J} {737, L45}

\bibitem[\protect\citeauthoryear{Kesden}{Kesden}{2012}]{kesden}
Kesden M.,  2012, \mn@doi [Phys. Rev. D] {10.1103/PhysRevD.85.024037}, 85, 024037

\bibitem[\protect\citeauthoryear{{Kippenhahn} \& {Weigert}}{{Kippenhahn} \& {Weigert}}{1990}]{kippen}
{Kippenhahn} R.,  {Weigert} A.,  1990, {Stellar Structure and Evolution}

\bibitem[\protect\citeauthoryear{Kormendy \& Ho}{Kormendy \& Ho}{2013}]{kormendy}
Kormendy J.,  Ho L.~C.,  2013, \mn@doi [Annual Review of Astronomy and Astrophysics] {https://doi.org/10.1146/annurev-astro-082708-101811}, 51, 511

\bibitem[\protect\citeauthoryear{{Kormendy} \& {Kennicutt}}{{Kormendy} \& {Kennicutt}}{2004}]{kormendy_n_kennicutt}
{Kormendy} J.,  {Kennicutt} Jr. R.~C.,  2004, \mn@doi [\araa] {10.1146/annurev.astro.42.053102.134024}, \href {https://ui.adsabs.harvard.edu/abs/2004ARA&A..42..603K} {42, 603}

\bibitem[\protect\citeauthoryear{{Krajnovi{\'c}} et~al.,}{{Krajnovi{\'c}} et~al.}{2018}]{krajnovi}
{Krajnovi{\'c}} D.,  et~al., 2018, \mn@doi [\mnras] {10.1093/mnras/sty778}, \href {https://ui.adsabs.harvard.edu/abs/2018MNRAS.477.3030K} {477, 3030}

\bibitem[\protect\citeauthoryear{Lacy \& Townes}{Lacy \& Townes}{1982}]{lacey}
Lacy J.~H.,  Townes C.~H.,  1982, \mn@doi [The Astrophysical Journal] {10.1086/160402}, 262, 120

\bibitem[\protect\citeauthoryear{Leja, Johnson, Conroy, van Dokkum  \& Byler}{Leja et~al.}{2017}]{leja}
Leja J.,  Johnson B.~D.,  Conroy C.,  van Dokkum P.~G.,   Byler N.,  2017, \mn@doi [The Astrophysical Journal] {https://doi.org/10.3847/1538-4357/aa5ffe}, 837, 170

\bibitem[\protect\citeauthoryear{Leja, Carnall, Johnson, Conroy  \& Speagle}{Leja et~al.}{2019}]{leja19}
Leja J.,  Carnall A.~C.,  Johnson B.~D.,  Conroy C.,   Speagle J.~S.,  2019, \mn@doi [The Astrophysical Journal] {10.3847/1538-4357/ab133c}, 876, 3

\bibitem[\protect\citeauthoryear{{Magorrian} et~al.,}{{Magorrian} et~al.}{1998}]{magorrian}
{Magorrian} J.,  et~al., 1998, \mn@doi [\aj] {10.1086/300353}, \href {https://ui.adsabs.harvard.edu/abs/1998AJ....115.2285M} {115, 2285}

\bibitem[\protect\citeauthoryear{{Martin} et~al.,}{{Martin} et~al.}{2005}]{galex}
{Martin} D.~C.,  et~al., 2005, \mn@doi [\apjl] {10.1086/426387}, \href {https://ui.adsabs.harvard.edu/abs/2005ApJ...619L...1M} {619, L1}

\bibitem[\protect\citeauthoryear{{McConnell} \& {Ma}}{{McConnell} \& {Ma}}{2013}]{mcconnell}
{McConnell} N.~J.,  {Ma} C.-P.,  2013, \mn@doi [\apj] {10.1088/0004-637X/764/2/184}, \href {https://ui.adsabs.harvard.edu/abs/2013ApJ...764..184M} {764, 184}

\bibitem[\protect\citeauthoryear{Mockler, Guillochon  \& Ramirez-Ruiz}{Mockler et~al.}{2019}]{mockler}
Mockler B.,  Guillochon J.,   Ramirez-Ruiz E.,  2019, \mn@doi [The Astrophysical Journal] {https://doi.org/10.3847/1538-4357/ab010f}, 872, 151

\bibitem[\protect\citeauthoryear{{Mummery}}{{Mummery}}{2024}]{mummery2023b}
{Mummery} A.,  2024, \mn@doi [\mnras] {10.1093/mnras/stad3636}, \href {https://ui.adsabs.harvard.edu/abs/2024MNRAS.527.6233M} {527, 6233}

\bibitem[\protect\citeauthoryear{Mummery \& Balbus}{Mummery \& Balbus}{2020}]{mummery20}
Mummery A.,  Balbus S.~A.,  2020, \mn@doi [Monthly Notices of the Royal Astronomical Society] {10.1093/mnras/staa192}, 492, 5655

\bibitem[\protect\citeauthoryear{{Mummery}, {van Velzen}, {Nathan}, {Ingram}, {Hammerstein}, {Fraser-Taliente}  \& {Balbus}}{{Mummery} et~al.}{2024}]{mummery2023}
{Mummery} A.,  {van Velzen} S.,  {Nathan} E.,  {Ingram} A.,  {Hammerstein} E.,  {Fraser-Taliente} L.,   {Balbus} S.,  2024, \mn@doi [\mnras] {10.1093/mnras/stad3001}, \href {https://ui.adsabs.harvard.edu/abs/2024MNRAS.527.2452M} {527, 2452}

\bibitem[\protect\citeauthoryear{{Muzzin} et~al.,}{{Muzzin} et~al.}{2013}]{muzzin}
{Muzzin} A.,  et~al., 2013, \mn@doi [\apj] {10.1088/0004-637X/777/1/18}, \href {https://ui.adsabs.harvard.edu/abs/2013ApJ...777...18M} {777, 18}

\bibitem[\protect\citeauthoryear{Nguyen et~al.,}{Nguyen et~al.}{2019}]{nguyen}
Nguyen D.~D.,  et~al., 2019, \mn@doi [The Astrophysical Journal] {10.3847/1538-4357/aafe7a}, 872, 104

\bibitem[\protect\citeauthoryear{{Nicholl}, {Lanning}, {Ramsden}, {Mockler}, {Lawrence}, {Short}  \& {Ridley}}{{Nicholl} et~al.}{2022}]{nicholl2022}
{Nicholl} M.,  {Lanning} D.,  {Ramsden} P.,  {Mockler} B.,  {Lawrence} A.,  {Short} P.,   {Ridley} E.~J.,  2022, arXiv e-prints, \href {https://ui.adsabs.harvard.edu/abs/2022arXiv220102649N} {p. arXiv:2201.02649}

\bibitem[\protect\citeauthoryear{{Phinney}}{{Phinney}}{1989}]{phinney}
{Phinney} E.~S.,  1989, in {Morris} M.,  ed.,  IAU Symposium Vol. 136, The Center of the Galaxy. p.~543

\bibitem[\protect\citeauthoryear{{Piran}, {Svirski}, {Krolik}, {Cheng}  \& {Shiokawa}}{{Piran} et~al.}{2015}]{piran}
{Piran} T.,  {Svirski} G.,  {Krolik} J.,  {Cheng} R.~M.,   {Shiokawa} H.,  2015, \mn@doi [ApJ] {10.1088/0004-637X/806/2/164}, \href {https://ui.adsabs.harvard.edu/abs/2015ApJ...806..164P} {806, 164}

\bibitem[\protect\citeauthoryear{Ramsden, Lanning, Nicholl  \& McGee}{Ramsden et~al.}{2022}]{ramsden}
Ramsden P.,  Lanning D.,  Nicholl M.,   McGee S.~L.,  2022, \mn@doi [Monthly Notices of the Royal Astronomical Society] {10.1093/mnras/stac1810}, 515, 1146

\bibitem[\protect\citeauthoryear{Rees}{Rees}{1988}]{rees}
Rees M.,  1988, \mn@doi [Nature] {https://doi.org/10.1038/333523a0}, 333, 523–528

\bibitem[\protect\citeauthoryear{Reines \& Volonteri}{Reines \& Volonteri}{2015}]{reines15}
Reines A.~E.,  Volonteri M.,  2015, \mn@doi [The Astrophysical Journal] {10.1088/0004-637x/813/2/82}, 813, 82

\bibitem[\protect\citeauthoryear{{Robotham}, {Taranu}, {Tobar}, {Moffett}  \& {Driver}}{{Robotham} et~al.}{2017}]{robotham2017}
{Robotham} A.~S.~G.,  {Taranu} D.~S.,  {Tobar} R.,  {Moffett} A.,   {Driver} S.~P.,  2017, \mn@doi [\mnras] {10.1093/mnras/stw3039}, \href {https://ui.adsabs.harvard.edu/abs/2017MNRAS.466.1513R} {466, 1513}

\bibitem[\protect\citeauthoryear{Roth, Kasen, Guillochon  \& Ramirez-Ruiz}{Roth et~al.}{2016}]{roth16}
Roth N.,  Kasen D.,  Guillochon J.,   Ramirez-Ruiz E.,  2016, \mn@doi [The Astrophysical Journal] {10.3847/0004-637x/827/1/3}, 827, 3

\bibitem[\protect\citeauthoryear{{Ryu}, {Krolik}  \& {Piran}}{{Ryu} et~al.}{2020}]{ryu}
{Ryu} T.,  {Krolik} J.,   {Piran} T.,  2020, \mn@doi [\apj] {10.3847/1538-4357/abbf4d}, \href {https://ui.adsabs.harvard.edu/abs/2020ApJ...904...73R} {904, 73}

\bibitem[\protect\citeauthoryear{{Sazonov} et~al.,}{{Sazonov} et~al.}{2021}]{sazonov}
{Sazonov} S.,  et~al., 2021, \mn@doi [\mnras] {10.1093/mnras/stab2843}, \href {https://ui.adsabs.harvard.edu/abs/2021MNRAS.508.3820S} {508, 3820}

\bibitem[\protect\citeauthoryear{{Schechter}}{{Schechter}}{1976}]{schechter}
{Schechter} P.,  1976, \mn@doi [\apj] {10.1086/154079}, \href {https://ui.adsabs.harvard.edu/abs/1976ApJ...203..297S} {203, 297}

\bibitem[\protect\citeauthoryear{Schlafly, Meisner  \& Green}{Schlafly et~al.}{2019}]{unwise}
Schlafly E.~F.,  Meisner A.~M.,   Green G.~M.,  2019, \mn@doi [The Astrophysical Journal Supplement Series] {10.3847/1538-4365/aafbea}, 240, 30

\bibitem[\protect\citeauthoryear{Schutte, Reines  \& Greene}{Schutte et~al.}{2019}]{schutte}
Schutte Z.,  Reines A.~E.,   Greene J.~E.,  2019, \mn@doi [The Astrophysical Journal] {10.3847/1538-4357/ab35dd}, 887, 245

\bibitem[\protect\citeauthoryear{{Skrutskie} et~al.,}{{Skrutskie} et~al.}{2006}]{2mass}
{Skrutskie} M.~F.,  et~al., 2006, \mn@doi [\aj] {10.1086/498708}, \href {https://ui.adsabs.harvard.edu/abs/2006AJ....131.1163S} {131, 1163}

\bibitem[\protect\citeauthoryear{Stone, Generozov, Vasiliev  \& Metzger}{Stone et~al.}{2018}]{stone18}
Stone N.~C.,  Generozov A.,  Vasiliev E.,   Metzger B.~D.,  2018, \mn@doi [Monthly Notices of the Royal Astronomical Society] {https://doi.org/10.1093/mnras/sty2045}, 480, 5060–5077

\bibitem[\protect\citeauthoryear{Thater, Krajnović, Cappellari, Davis, de Zeeuw, McDermid  \& Sarzi}{Thater et~al.}{2019}]{thater}
Thater S.,  Krajnović D.,  Cappellari M.,  Davis T.~A.,  de Zeeuw P.~T.,  McDermid R.~M.,   Sarzi M.,  2019, \mn@doi [Astronomy &amp; Astrophysics] {10.1051/0004-6361/201834808}, 625, A62

\bibitem[\protect\citeauthoryear{{Tremaine} et~al.,}{{Tremaine} et~al.}{2002}]{tremaine}
{Tremaine} S.,  et~al., 2002, \mn@doi [\apj] {10.1086/341002}, \href {https://ui.adsabs.harvard.edu/abs/2002ApJ...574..740T} {574, 740}

\bibitem[\protect\citeauthoryear{Wevers, Stone, van Velzen, Jonker, Hung, Auchettl, Gezari  \& Onori}{Wevers et~al.}{2019}]{wevers19}
Wevers T.,  Stone N.~C.,  van Velzen S.,  Jonker P.~G.,  Hung T.,  Auchettl K.,  Gezari S.,   Onori F.,  2019, \mn@doi [Monthly Notices of the Royal Astronomical Society] {10.1093/mnras/stz1602}, 487, 4136

\bibitem[\protect\citeauthoryear{{Wright} et~al.,}{{Wright} et~al.}{2010}]{wright}
{Wright} E.~L.,  et~al., 2010, \mn@doi [\aj] {10.1088/0004-6256/140/6/1868}, \href {https://ui.adsabs.harvard.edu/abs/2010AJ....140.1868W} {140, 1868}

\bibitem[\protect\citeauthoryear{Yao et~al.,}{Yao et~al.}{2023}]{yao23}
Yao Y.,  et~al., 2023, Tidal Disruption Event Demographics with the Zwicky Transient Facility: Volumetric Rates, Luminosity Function, and Implications for the Local Black Hole Mass Function (\mn@eprint {arXiv} {2303.06523}), \url {https://arxiv.org/abs/2303.06523}

\bibitem[\protect\citeauthoryear{{York} et~al.,}{{York} et~al.}{2000}]{sdss}
{York} D.~G.,  et~al., 2000, \mn@doi [\aj] {10.1086/301513}, \href {https://ui.adsabs.harvard.edu/abs/2000AJ....120.1579Y} {120, 1579}

\bibitem[\protect\citeauthoryear{van Velzen}{van Velzen}{2018}]{vanvelzen2018}
van Velzen S.,  2018, \mn@doi [The Astrophysical Journal] {https://doi.org/10.3847/1538-4357/aa998e}, 852, 72

\bibitem[\protect\citeauthoryear{{van Velzen} et~al.,}{{van Velzen} et~al.}{2011}]{van_class5}
{van Velzen} S.,  et~al., 2011, \mn@doi [The Astrophysical Journal] {10.1088/0004-637X/741/2/73}, \href {https://ui.adsabs.harvard.edu/abs/2011ApJ...741...73V} {741, 73}

\bibitem[\protect\citeauthoryear{{van Velzen}, {Holoien}, {Onori}, {Hung}  \& {Arcavi}}{{van Velzen} et~al.}{2020}]{vanvelzen2020}
{van Velzen} S.,  {Holoien} T. W.~S.,  {Onori} F.,  {Hung} T.,   {Arcavi} I.,  2020, \mn@doi [\ssr] {10.1007/s11214-020-00753-z}, \href {https://ui.adsabs.harvard.edu/abs/2020SSRv..216..124V} {216, 124}

\makeatother
\end{thebibliography}





\appendix

\section{The BH--Total Mass Relation}
\label{a:A}

To assess the impact of our choice of host galaxy bulge versus total stellar mass in our analysis, the Prospector fitting procedure described in Section \ref{seds} was used to recompute total stellar masses for the \cite{kormendy} sample. This also ensures consistency in modelling across samples, particularly in terms of star formation history and stellar population assumptions, which can significantly influence stellar mass estimates \citep{schutte}. Note that as in the bulge mass sample, we exclude any data that \cite{kormendy} omitted in fitting as well as any sources identified as ongoing mergers. Additionally, targets were automatically removed if they had: very small redshifts, such that the projected size of the galaxy was larger than the apertures used for survey photometry; inadequate broadband photometry; or could not be assigned sky coordinates. Five further targets were omitted after fitting due to unphysically low total stellar mass estimates, indicating an issue with the available photometry. Following these cuts, the final total stellar mass sample consists of 52 high mass galaxies. Analysis was repeated, as in section \ref{data-analysis}, for the SMBH--Total mass relation using this TDE+KH13 total stellar mass sample.

\begin{figure*}
\centering
\includegraphics[width=1\textwidth]{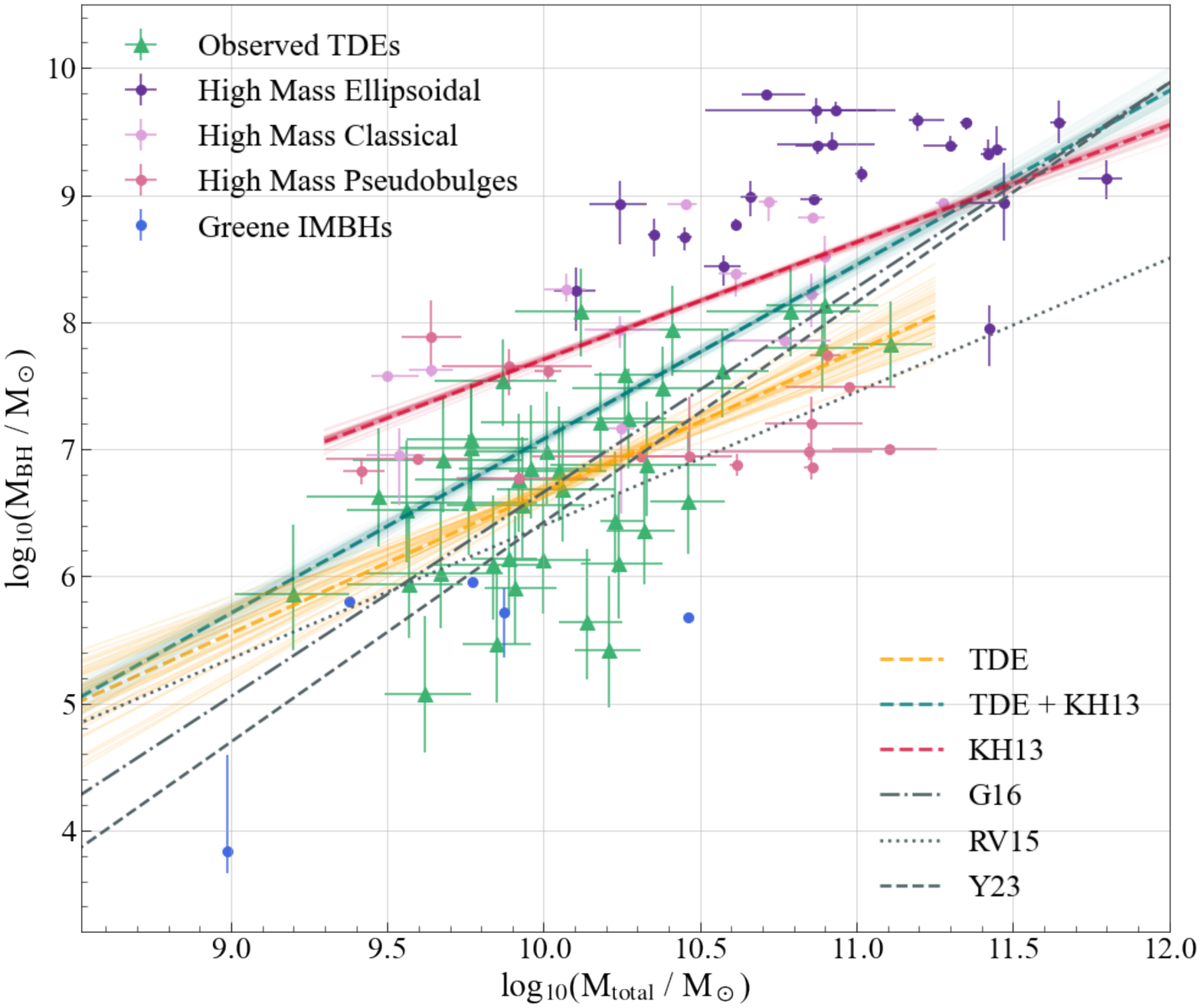}
\caption{SMBH mass as a function of host galaxy total stellar mass for the TDE sample (green, showing statistical errors only) and the high-mass regime from Kormendy \& Ho (2013) re-computed using Prospector fitting. This includes ellipsoidal galaxies (dark purple) and classical bulges (light purple), plus pseudobulges (pink), which were excluded from the original KH13 fit. The dashed teal line, Eq \ref{comb_total_fit}, shows the average fit to the TDE total stellar mass sample after bootstrapping with perturbation, and the dashed orange and red lines show the equivalent for the TDE only and KH13 only samples respectively, Eq \ref{tde_total_fit} and Eq \ref{hmr_total_fit}. The shaded regions show the spread of bootstrapping with perturbation fits. A sample of non-TDE low mass galaxies with nuclear BH detections from the Greene et al. (2020) IMBH review have been plotted but not fit (blue). Additional fits to the black hole-stellar mass relation that include low mass galaxies have also been overlayed, including relations from Greene et al. (2020), Reines \& Volonteri (2015), and Yao et al. (2023). \label{big_total_scatter}}
\end{figure*}

The power-law relation for the TDE total stellar mass sample has a gradient of $1.11^{+0.13}_{-0.12}$ and a y-intercept of $-4.431^{+1.22}_{-1.26}$, with a Spearman coefficient $r_s = 0.51^{+0.06}_{-0.06}$, $P_s < 0.001$. This gives a SMBH-Total mass relationship: 

\begin{equation}\label{tde_total_fit}
\frac{M_{\text{BH}}}{10^9M_{\odot}} = (0.06^{+0.02}_{-0.02}) \cdot \left (\frac{M_{\rm bulge}}{10^{11}M_{\odot}}\right) ^{1.11^{\pm 0.12}}; \qquad \epsilon_0 = 0.41 \text{ dex}. 
\end{equation}

Comparatively, for the TDE+KH13 sample we find a Spearman coefficient $r_s = 0.68^{+0.02}_{-0.02}$, with significance $P_s < 0.001$. The best-fit straight line has a gradient of $1.37^{+0.05}_{-0.04}$ and a y-intercept of $-6.63^{+0.46}_{-0.46}$, which can be expressed as:

\begin{equation}\label{comb_total_fit}
\frac{M_{\text{BH}}}{10^9M_{\odot}} = (0.28^{+0.02}_{-0.02}) \cdot \left (\frac{M_{\rm bulge}}{10^{11}M_{\odot}}\right) ^{1.37\pm 0.04}; \qquad \epsilon_0 = 0.78 \text{ dex}.  
\end{equation}

In order to make comparisons between the TDE+KH13 sample and that of the high mass regime, we apply the bootstrapping with perturbation described in section \ref{data-analysis} to the \citep{kormendy} total stellar mass sample. We find a power-law relation with gradient $0.92^{+0.03}_{-0.03}$ and y-intercept $-1.51^{+0.33}_{-0.34}$, with a Spearman coefficient $r_s = 0.57^{+0.02}_{-0.03}$, $P_s < 0.001$. This gives a SMBH-Total mass relationship of:

\begin{equation}\label{hmr_total_fit}
\frac{M_{\text{BH}}}{10^9M_{\odot}} = (0.43^{+0.09}_{-0.10}) \cdot \left (\frac{M_{\rm bulge}}{10^{11}M_{\odot}}\right) ^{0.92\pm 0.03}; \qquad \epsilon_0 = 0.77 \text{ dex}.  
\end{equation}

Alongside the derived total stellar mass relations, figure \ref{big_total_scatter} highlights the robustness of our main results to the choice of stellar mass definition. Notably, the relation derived for the TDE-only sample remains statistically significant and comparatively flatter than that of the TDE+KH13 sample in the total stellar mass regime. Additionally, the TDE+KH13 sample continues to support the need for a steeper slope than previously expected from canonical relationships, this is seen when comparing the TDE+KH13 total stellar mass sample gradient to that derived for the \cite{kormendy} total stellar mass sample. However, compared to the bulge mass relations, relationships based on total stellar mass exhibit significantly greater scatter. Hence, while this total stellar mass analysis is valuable in testing our results, we continue to use bulge mass in the main analysis due to its tighter correlation.

A small sample of non-TDE low-mass galaxies with nuclear BH detections from \cite{greene20} is shown in the context of the derived total stellar mass relations. This IMBH sample excludes any known TDE hosts that are already included in our sample and is not considered in the analysis as we adopt the published stellar masses and do not perform Prospector fits to the data. Nonetheless, from fig \ref{big_total_scatter} it is clear that these lower mass galaxies are largely consistent with the TDE hosts, overlapping in the $M_{\text{BH}} > 10^5$ regime. With a single data point, the IMBH sample extends the relationship down to $M_{\text{BH}} \sim 10^4$. While only one such system exists in this mass range, it falls within the same trend as the TDE hosts consistent with a simple extrapolation. This supports the position of TDEs on this SMBH--Total mass relation as well as reinforcing the need for a steeper slope if lower mass galaxies and their BHs are to be included in the relationship.

\begin{table}
\small
\caption{Power-law fits to SMBH--Total stellar mass from this work (TDE+KH13; KH13), Reines \& Volenteri 2015 (RV15), Greene et al. 2020 (G20) and Yao et al. 2023 (Y23).}
\label{total-powerlaw-fits}
        \begin{center}
\renewcommand{\arraystretch}{1.6}
\begin{tabular}{l l l l} 
 \hline\hline
  & $\alpha$ & $\beta$ & $\epsilon_0$  \\ 
 \hline
 TDE+KH13 & $0.2800 (\pm0.0200)$ & $1.3700 (\pm0.0400)$
  & $0.78$ \\
 KH13 &  $0.4300(\pm0.0900)$ & $0.9200 (\pm0.0300)$ & $0.77$ \\
 RV15 & $0.0280(\pm0.0003)$ & $1.0500(\pm0.1100)$ & $0.55$ \\
 G20 & $0.1900(\pm0.0020)$ & $1.6100(\pm0.1200)$ & $0.81$ \\
 Y23 & $1.4200(\pm0.1100)\cdot 10^{-11}$ & $1.7300 (\pm0.2300) $ & $0.17$ \\
  \hline\hline
\end{tabular}
\hypersetup{citecolor=blue}
\end{center}
\small \textbf{Note:} All SMBH--Total mass relations are of the form, \\ ($M_{\rm BH}/10^{11}M_{\odot}) = \alpha (M_{\rm bulge}/10^9 M_{\odot})^{\beta}$ and $\epsilon_0$ is the intrinsic scatter.
\end{table}

Additionally, we over-plot a selection of literature fits to the BH–total stellar mass relation that extend into the low-mass regime \citep{reines15,greene20,yao23}. These studies include galaxies with BH masses ranging from $10^3 M_{\odot}$ to $10^{10} M_{\odot}$ \citep{greene20}, overlapping significantly with the mass range explored in this work. The relations derived by \cite{greene20} and \cite{yao23} have notably similar slopes, with gradients of $1.61\pm0.12$ and $1.73\pm0.23$ respectively. Variations between the two may be due to the narrower BH mass range seen in \cite{yao23} ($10^{5.0} M_{\odot} \lesssim M_{\text{BH}} \lesssim 10^{8.5} M_{\odot}$) as well as differences in initial mass functions and population assumptions. In comparison, the \cite{reines15} relation is flatter than both our TDE+KH13 result and the other literature fits, with a gradient of $1.05\pm0.11$. This likely results from a combination of the large systematic uncertainties associated with broad-line AGN estimates; a narrower BH mass range; and differences in stellar mass estimation methods. which could partially explain the shallower slope.

Overall, accounting for differences in methods of stellar mass estimation, population assumptions and sample sizes, our findings are consistent with those of prior studies. All literature relations over-plotted exhibit steeper slopes than that derived for the \cite{kormendy} total stellar mass sample. This is consistent with our conclusion that a steepening is required to accurately describe the BH--Total mass relation when extended to the low-mass regime.


\bsp	
\label{lastpage}
\end{document}